\documentclass[preprintnumbers,article,amsmath,amssymb,floatfix,10pt,prd,superscriptaddress,nofootinbib]{revtex4}

\usepackage{doi}
\usepackage{hyperref}
\hypersetup{
  colorlinks=true,       
  linkcolor=blue,         
  citecolor=cyan,         
}

\usepackage{graphicx}% Include figure files
\usepackage{dcolumn}% Align table columns on decimal point
\usepackage{bm}% bold math
\usepackage{epstopdf}
\usepackage{color}
\usepackage{enumitem}

\newcommand{\beq}{\begin{equation}}
\newcommand{\eeq}{\end{equation}}
\newcommand{\bea}{\begin{eqnarray}}
\newcommand{\eea}{\end{eqnarray}}

\usepackage{bbm}
\usepackage{amsfonts}
\usepackage{mathrsfs}
\usepackage{latexsym}
\usepackage{epsfig}
\usepackage{epstopdf}
\usepackage{dcolumn}
\usepackage{bm}
\usepackage{color}
\usepackage{comment}
\usepackage{xcolor}
\usepackage{orcidlink}

\begin{document}
\title{Energy extraction processes and thermodynamics in rotating Weyl conformal black holes}

\author{Dilmurod~Ortiqboev}
\email{artdima93@gmail.com}
\affiliation{Samarkand State University, University Avenue 15, 140104 Samarkand, Uzbekistan}
\affiliation{Gulistan State University, Gulistan 120100, Uzbekistan}

\author{Farruh~Atamurotov\orcidlink{0000-0001-8857-4970}}
\email{atamurotov@yahoo.com}
\affiliation{Research Center of Astrophysics and Cosmology, Khazar University, 41 Mehseti Street, Baku AZ1096, Azerbaijan}
\affiliation{Kimyo International University in Tashkent, Shota Rustaveli str. 156, Tashkent 100121, Uzbekistan}
\affiliation{University of Tashkent for Applied Sciences, Str.Gavhar 1, Tashkent 100149, Uzbekistan}

\author{Ahmadjon~Abdujabbarov}
\email{ahmadjon@astrin.uz}
\affiliation{School of Physics, Harbin Institute of Technology, Harbin 150001, People’s Republic of China}
\affiliation{Institute of Fundamental and Applied Research, National Research University TIIAME, Kori Niyoziy 39, Tashkent 100000, Uzbekistan}
\affiliation{National University of Uzbekistan, Tashkent 100174, Uzbekistan}

\author{Chen Zhou}
\email{chenzhou@hit.edu.cn (Corresponding Author)}
\affiliation{School of Physics, Harbin Institute of Technology, Harbin 150001, People’s Republic of China}

\author{G. Mustafa\orcidlink{0000-0003-1409-2009}}
\email{gmustafa3828@gmail.com (Corresponding Author)}
\affiliation{Department of Physics, Zhejiang Normal University, Jinhua 321004, China}

\begin{abstract}
This work investigates the original Penrose process (PP), as well as black hole (BH) thermodynamics, within the framework of conformal Weyl gravity. The theory is characterized by the dimensionful constants $\gamma$, which encodes the asymptotic rotational curvature of galactic rotation curves in an otherwise asymptotically flat spacetime, and $\kappa$, which is associated with the cosmological background constant $\Lambda$.  We determine the locations of the event horizon and the extreme configurations for rotating BHs in this theory, and we provide both perturbative and exact expressions for the event horizon radius. The geodesic equations are restricted to the equatorial plane, and the corresponding dynamics, including energy extraction processes, are studied exclusively in this plane. As the principal results, we examine the feasibility and efficiency of the original  PP, analyze BH deceleration mechanisms, and test the validity of the four laws of BH thermodynamics in the context of conformal Weyl gravity.

\end{abstract}

\maketitle

\date{\today}

%\tableofcontents

%%%%%%%%%%%%%%%%%%%%%%%%%%%%%%%%%%%%%%%%%%%%%%%%%
\section{Introduction}\label{S1}
%%%%%%%%%%%%%%%%%%%%%%%%%%%%%%%%%%%%%%%%%%%%%%%%%

General Relativity (GR), formulated by Albert Einstein in 1915 \cite{Einstein1915}, remains the most successful classical theory of gravitation and provides the fundamental framework for describing gravitational phenomena on astrophysical and cosmological scales. In this theory, gravity is interpreted not as a force but as the manifestation of spacetime curvature generated by matter and energy through the Einstein field equations. GR has achieved remarkable agreement with observational and experimental tests, including the perihelion precession of Mercury, gravitational lensing, BH physics, and the recent direct detection of gravitational waves \cite{Abbott2016}. Moreover, BH solutions such as the Schwarzschild \cite{Schwarzschild1916} and Kerr geometries \cite{Kerr1963} play a central role in modern relativistic astrophysics and high-energy physics, offering important insights into strong gravitational fields, particle dynamics, and quantum aspects of gravity. Despite its tremendous success, several unresolved problems, including the dark matter and dark energy puzzles as well as the non-renormalizability of gravity at the quantum level, motivate the study of alternative and modified theories of gravity beyond Einstein’s framework .

Weyl conformal gravity is one of the most extensively studied higher-order alternatives to General Relativity, originally proposed by Hermann Weyl as a locally conformal invariant theory of gravitation \cite{Weyl1918}. In contrast to Einstein gravity, conformal Weyl gravity is invariant under local conformal transformations of the metric \cite{Mannheim2006,Flanagan2006}. The corresponding gravitational action is uniquely constructed from the square of the Weyl tensor and leads to fourth-order field equations governed by the Bach tensor \cite{Bach1921,Mannheim1989}. One of the remarkable features of Weyl gravity is that it naturally provides modifications to the gravitational potential at galactic and cosmological scales without introducing dark matter, successfully explaining flat galactic rotation curves through additional linear and quadratic terms in the metric function \cite{Mannheim1989,Mannheim2012,Varieschi2010}. Moreover, conformal gravity has attracted considerable attention in BH physics, cosmology, and quantum gravity due to its renormalizable structure and richer geometrical properties compared to Einstein’s theory \cite{tHooft2015,Riegert1984}. In recent years, rotating BH solutions in conformal Weyl gravity and their astrophysical implications have been actively investigated, including particle dynamics, geodesic motion, thermodynamics, and energy extraction processes around Kerr-like Weyl BHs \cite{Asuncion2025,Turner2020}.

The original PP, proposed by Roger Penrose in 1969, is one of the most important mechanisms for extracting rotational energy from rotating BHs \cite{Penrose1969}. The process occurs inside the ergoregion of a rotating BH, where the dragging of inertial frames allows particles with negative energy, to exist \cite{Misner1973}. In the classical scenario, an incoming particle enters the ergosphere and splits into two fragments: one particle falls into the BH with negative energy, while the second escapes to large distance carrying more energy than the initial particle, thereby extracting rotational energy from the BH \cite{Bardeen1972,Chandrasekhar1983}. This mechanism has been extensively investigated in Kerr geometry and later generalized to various modified theories of gravity and non-Kerr space-times \cite{Wagh1985,Brito2015}. In the context of conformal Weyl gravity, the PP has recently attracted renewed interest due to the modified horizon structure and ergo-region produced by conformal corrections \cite{Asuncion2025}. Studies of the original Penrose scenario have been carried out for static and rotating BH solutions in Mannheim--Kazanas conformal gravity, Kerr-like Weyl BHs, and Einstein--Weyl higher-derivative gravity models \cite{Mannheim1989,Turner2020,Asuncion2025}. These investigations demonstrate that the conformal parameters can significantly modify the efficiency of energy extraction, the structure of circular orbits, and the thermodynamic properties of rotating BHs \cite{Mannheim2006,Asuncion2025}.

BH thermodynamics establishes a profound connection between gravitation, quantum theory, and statistical physics by attributing thermodynamic properties such as temperature and entropy to BHs \cite{Bekenstein1973, Hawking1975}. The four laws of BH mechanics were formulated in close analogy with the laws of ordinary thermodynamics, where the surface gravity corresponds to temperature and the horizon area corresponds to entropy \cite{Bardeen1973}. Hawking’s discovery of BH radiation demonstrated that BHs are thermal objects with temperature, while the Bekenstein--Hawking entropy is proportional to the area of the event horizon  \cite{Hawking1975, Bekenstein1973}. In modified theories of gravity, especially higher-derivative theories, the entropy generally deviates from the simple area law and must instead be computed using the Wald Noether-charge formalism \cite{Wald1993,Iyer1994}. Conformal Weyl gravity provides an important framework for studying such corrections because its action contains quadratic curvature terms and possesses local conformal symmetry \cite{Mannheim2006}. BH thermodynamics in conformal Weyl gravity has been investigated in several contexts, including static Mannheim--Kazanas BHs, asymptotically AdS conformal BHs, Einstein--Weyl gravity, and rotating Kerr-like Weyl solutions \cite{Riegert1984,Lu2012,Klemm1998,Lu2011,Asuncion2025}. These studies examined the validity of the four laws of thermodynamics, Hawking temperature, Wald entropy, conserved charges, phase transitions, and critical behavior of conformal BHs \cite{Maldacena2011,Lu2012,Asuncion2025}. In particular, higher-derivative contributions in conformal gravity modify the entropy, heat capacity, and thermodynamic stability of BHs, leading to richer phase structures compared to standard Kerr or Schwarzschild spacetimes \cite{Lu2011,Mannheim2006}.

The paper is organized as follows. In Sec.\ref{sec: Rotating Kerr BH in Weyl gravity}, we briefly review the rotating BH solution in conformal Weyl gravity and discuss the corresponding spacetime geometry, horizon structure, and basic properties of the metric. In Sec.\ref{sec: Equatorial equation of motion}, we investigate the motion of test particles in the equatorial plane and derive the effective potential governing the radial dynamics. Sec.\ref{sec: Extractable energy and PP} is devoted to the analysis of the original PP, including the extractable energy, energy efficiency, and BH spin-down mechanism in conformal Weyl gravity. In Sec.\ref{sec: BH thermodynamics}, we study the thermodynamic properties of rotating Weyl BHs and examine the validity of the four laws of BH thermodynamics, together with the corresponding entropy, Hawking temperature, conserved charges, and phase transitions. Finally, in Sec.\ref{sec:conclusion}, we summarize the main results and discuss possible astrophysical implications of the obtained solutions.

Throughout this paper, we use the natural system of units in which the gravitational constant, the reduced Planck constant, and the speed of light are set equal to unity, $G=\hbar=c=1$. In addition, we adopt the metric signature $(-,+,+,+)$.
%%%%%%%%%%%%%%%%%%%%%%%%%%%%%%%%%%%%%%%%%%%%%%%%%%%
\section{Rotating Kerr BH in Weyl gravity}\label{sec: Rotating Kerr BH in Weyl gravity}
%%%%%%%%%%%%%%%%%%%%%%%%%%%%%%%%%%%%%%%%%%%%%%%%%%%

Weyl conformal gravity \cite{Weyl:1918ib} constitutes a well-studied alternative to general relativity. While the Einstein field equations are invariant under general coordinate transformations $g_{\mu\nu}(x) \rightarrow g'_{\mu\nu}(x')$ and Lorentz transformations $x^\mu \rightarrow \Lambda^\mu_{\nu} x^\nu$, conformal gravity is additionally constructed to be invariant under local conformal rescalings of the metric,
\begin{equation}
g_{\mu\nu}(x) \rightarrow \tilde{g}_{\mu\nu}(x) = \Omega^2(x)\, g_{\mu\nu}(x),
\end{equation}
where $\Omega(x)$ denotes the conformal factor specifying the local scaling of the line element \cite{Turner:2020gxo, Varieschi:2009vlp}. The requirement of local conformal invariance for the gravitational field equations uniquely leads, at the level of fourth-order derivative theories, to an action functional of the form
\begin{equation}\label{eq1}
A_{\mathrm{Weyl}} = -\alpha_{\mathrm{g}} \int \mathrm{d}^4 x\, \sqrt{-g}\,C_{\mu\nu\rho\sigma} C^{\mu\nu\rho\sigma},
\end{equation}
where $\alpha_{\mathrm{g}}$ is a dimensionless gravitational coupling constant, $g$ is the determinant of the metric tensor $g_{\mu\nu}$, and $C_{\mu\nu\rho\sigma}$ is the Weyl tensor. The latter is defined as the completely traceless part of the Riemann curvature tensor $R_{\mu\nu\rho\sigma}$, obtained by subtracting all trace contributions constructed from the Ricci tensor and the Ricci scalar \cite{Alestas:2019wtw}.

\begin{eqnarray}
C_{\mu\nu\rho\sigma}=R_{\mu\nu\rho\sigma}- \frac{1}{2}\left(g_{\mu\rho} R_{\nu\sigma}-g_{\mu\sigma} R_{\nu\rho}+g_{\nu\sigma} R_{\mu\rho}-g_{\nu\rho} R_{\mu\sigma}\right)+\frac{1}{6} R\left(g_{\mu\rho} g_{\nu\sigma}-g_{\mu\sigma} g_{\nu\rho}\right) \label{Weyl tensor}
\end{eqnarray}
where, as usual, $R_{\mu\nu}=g^{\rho\sigma} R_{\rho\mu\sigma\nu}$ denotes the Ricci tensor. By varying the gravitational action $A_{\mathrm{Weyl}}$ from Eq. (\ref{eq1}) with respect to the metric $g_{\mu\nu}$, one obtains \cite{Mannheim:2005bfa}
\begin{equation}\label{eq3}
\frac{1}{\sqrt{-g}} \frac{\delta A_{\mathrm{Weyl}}}{\delta g_{\mu\nu}}=-2 \alpha_{\mathrm{g}} \mathcal{W}^{\mu\nu},
\end{equation}
where $\mathcal{W}^{\mu\nu}$ denotes the traceless Bach tensor. Equivalently, this can be expressed as
\begin{equation}\label{eq4}
\mathcal{W}_{\mu\nu}=\mathcal{W}_{\mu\nu}^{(b)}-\frac{1}{3} \mathcal{W}_{\mu\nu}^{(a)},
\end{equation}
with
\begin{eqnarray*}
\left\{\begin{array}{l}
\mathcal{W}_{\mu\nu}^{(a)}=2 g_{\mu\nu} R_{; \lambda}^{; \lambda}-2 R_{;\mu;\nu}-2 R R_{\mu\nu}+\frac{1}{2} R^2, \\
\mathcal{W}_{\mu\nu}^{(b)}=\frac{1}{2} g_{\mu\nu} R_{; \lambda}^{; \lambda}+R_{\mu\nu}^{; \lambda}{ }_{; \lambda}-R_\mu^\lambda{ }_{; \nu ; \lambda}-2 R_{\mu \lambda} R_\nu{ }^\lambda+\frac{1}{2} g_{\mu\nu} R_{\rho \lambda} R^{\rho \lambda}
\end{array}\right.
\end{eqnarray*}
This, in turn, permits the conformal gravity field equations to be cast in a representation that is formally analogous to the corresponding field equations of general relativity.
\begin{equation}\label{eq5}
\mathcal{W}_{\mu\nu}=\frac{1}{4\alpha_{\mathrm{g}}} T_{\mu\nu}
\end{equation}
Several fundamental deviations of conformal gravity from General Relativity (GR) become apparent upon analysis of the static, uncharged, spherically symmetric vacuum solution obtained by Mannheim and Kazanas \cite{Alestas:2019wtw}. We refer to this spacetime metric, which may be regarded as the conformal gravity analogue of the Schwarzschild solution, as
\begin{equation}
\mathrm{d} s^2=-D(r) \mathrm{d} t^2+\frac{\mathrm{d} r^2}{D(r)}+r^2\left(\mathrm{~d} \theta^2+\sin ^2 \theta \mathrm{~d} \phi^2\right)  \label{eq:static solution}
\end{equation}
The lapse function in the form
\begin{equation}
D(r)=1-3\gamma M-\frac{2\mathcal{M}}{r}+\gamma r-\kappa r^2,  \label{eq:lapce function}
\end{equation}
where $M$ is the BH mass parameter and $\mathcal{M}=(2-3\gamma M)M/2$ is called the effective mass. Here $\gamma$ helps to explain the rotation curve of galaxies in a flat space at infinity without dark matter, and on a galactic scale it is of the order of $\gamma \sim 10^{-26} \;\text{m}^{-1}$. When $\gamma=0$, $\kappa$ is cosmologically dominant with $\kappa=\Lambda/3 $, leading to the de-Sitter ($\kappa>0$) and anti-de-Sitter ($\kappa<0$) models, and on a cosmological scale it is of the order of $\kappa\sim \,10^{-52}\; \text{m}^{-2}$. By employing the procedure described in Ref.~\cite{Asuncion:2025cxw}, one can derive the rotating Kerr BH solution in Weyl gravity in the following form:
\begin{eqnarray}
\mathrm{d} s^2 &=&-\frac{\Delta^H-\Delta^\theta a^2 \sin^2\theta}{\rho^2} \mathrm{d} t^2 -2 \frac{\left(r^2+a^2\right)\Delta^\theta-\Delta^H}{\rho^2} a \sin^2\theta ~ \mathrm{d} t \mathrm{~d} \phi +\frac{\rho^2}{\Delta^{\mathrm{H}}} \mathrm{d} r^2+\frac{\rho^2}{\Delta^\theta} \mathrm{d} \theta^2+\frac{\Sigma^2 \sin ^2 \theta}{\rho^2} \mathrm{d} \phi^2 \label{rotating metric}
\end{eqnarray}
with the following auxiliary terms, which are defined as:
\begin{eqnarray*}
 k &\equiv & \kappa+\frac{\gamma^2(1-\gamma M)}{(2-3 \gamma M)^2},\;\;\;
 \rho^2 \equiv  r^2+a^2 \cos ^2 \theta,\;\;\;
 \Delta^{\mathrm{H}} \equiv -k r^4+r^2-2 \mathcal{M} r+a^2,\;\;\;
 \Delta^\theta \equiv  1-k a^2 \cos ^2 \theta \cot ^2 \theta \\
\Sigma^2 &\equiv &\Delta^\theta\left(r^2+a^2\right)^2-a^2 \Delta^{\mathrm{H}} \sin ^2 \theta,
\end{eqnarray*}
Assuming that the event horizon is described by a hypersurface of constant radius, $f(r)=r-r_H=0$, the associated normal co-vector is given by $n_\mu=\partial_\mu f=(0,1,0,0)$. The causal character of this hypersurface is determined by the norm of the normal, $n^\mu n_\mu = g^{\mu\nu}n_\mu n_\nu = g^{rr}$. For a stationary and axi-symmetric BH spacetime with inverse metric component $g^{rr}=\Delta^H/\rho^2$, the condition that the horizon be a null hypersurface requires $n^\mu n_\mu = \Delta^H/\rho^2=0$.
This implies that the horizon is located at the real roots of $\Delta^H(r)=0$, which generically yields four main solutions. In this work, we consider some energetic processes near of the event horizon, at this scale cosmological effects are small enough and the event horizon can be considered a first order correction of $kM^2$ as
\begin{align}
    r_+=\left(\mathcal{M}+\sqrt{\mathcal{M}^2-a^2}\right)\left(1+\frac{k\left(\mathcal{M}+\sqrt{\mathcal{M}^2-a^2}\right)^3}{2\sqrt{\mathcal{M}^2-a^2}}\right),\label{eq: outer horizon} \;\\
    r_-=\left(\mathcal{M}-\sqrt{\mathcal{M}^2-a^2}\right)\left(1-\frac{k\left(\mathcal{M}-\sqrt{\mathcal{M}^2-a^2}\right)^3}{2\sqrt{\mathcal{M}^2-a^2}}\right). \label{eq: inner horizon}
\end{align}
\begin{figure}
    \centering
    \includegraphics[scale=0.4]{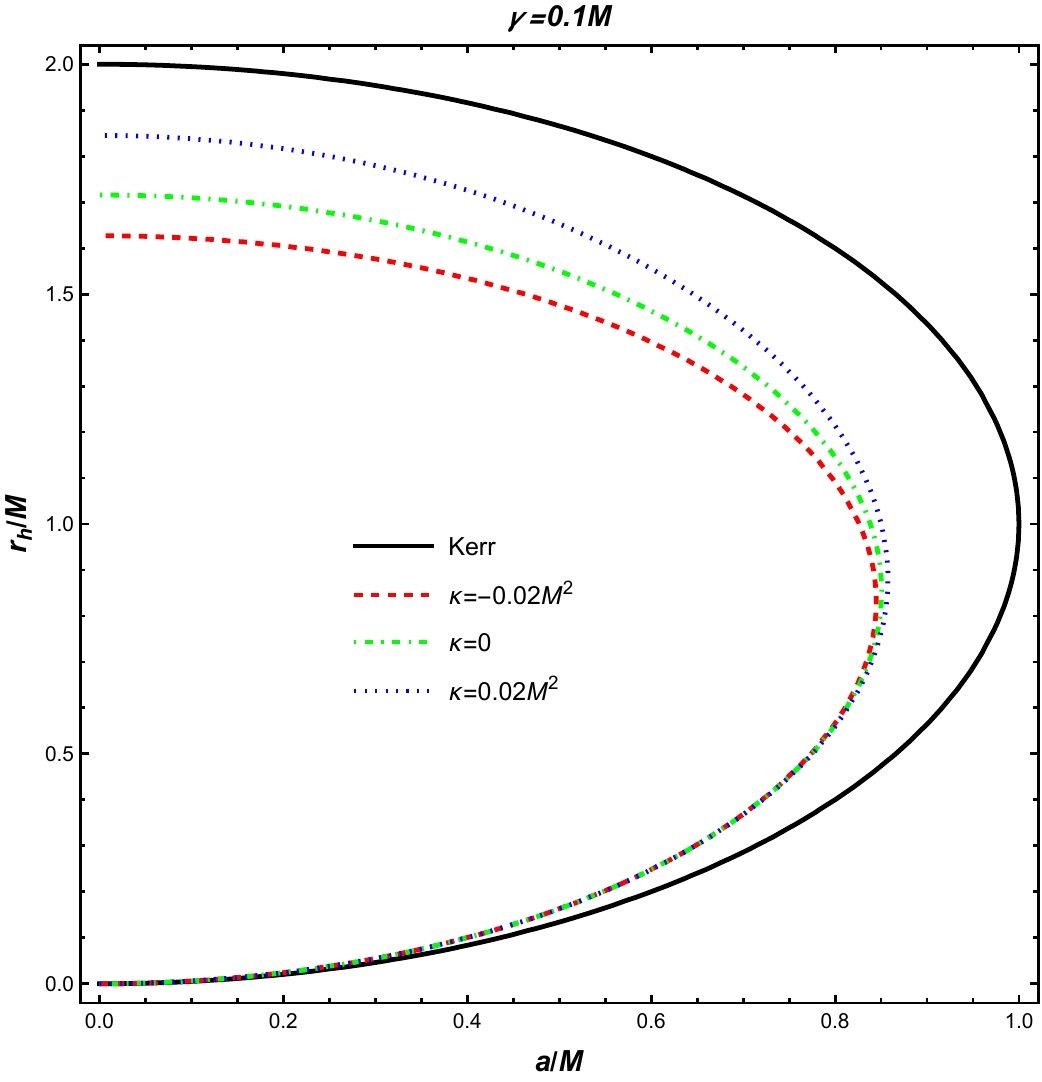}
    \includegraphics[scale=0.4]{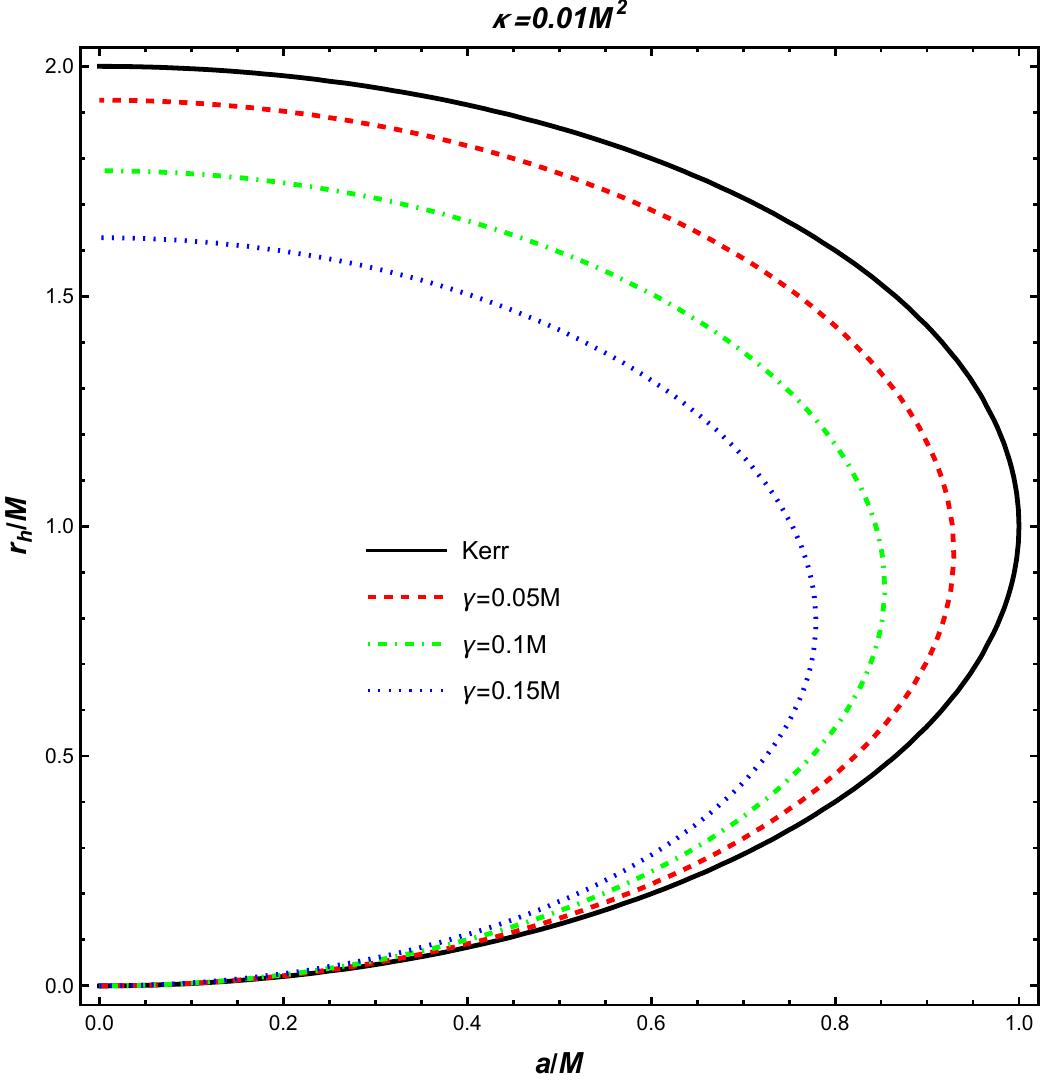}
    \includegraphics[scale=0.4]{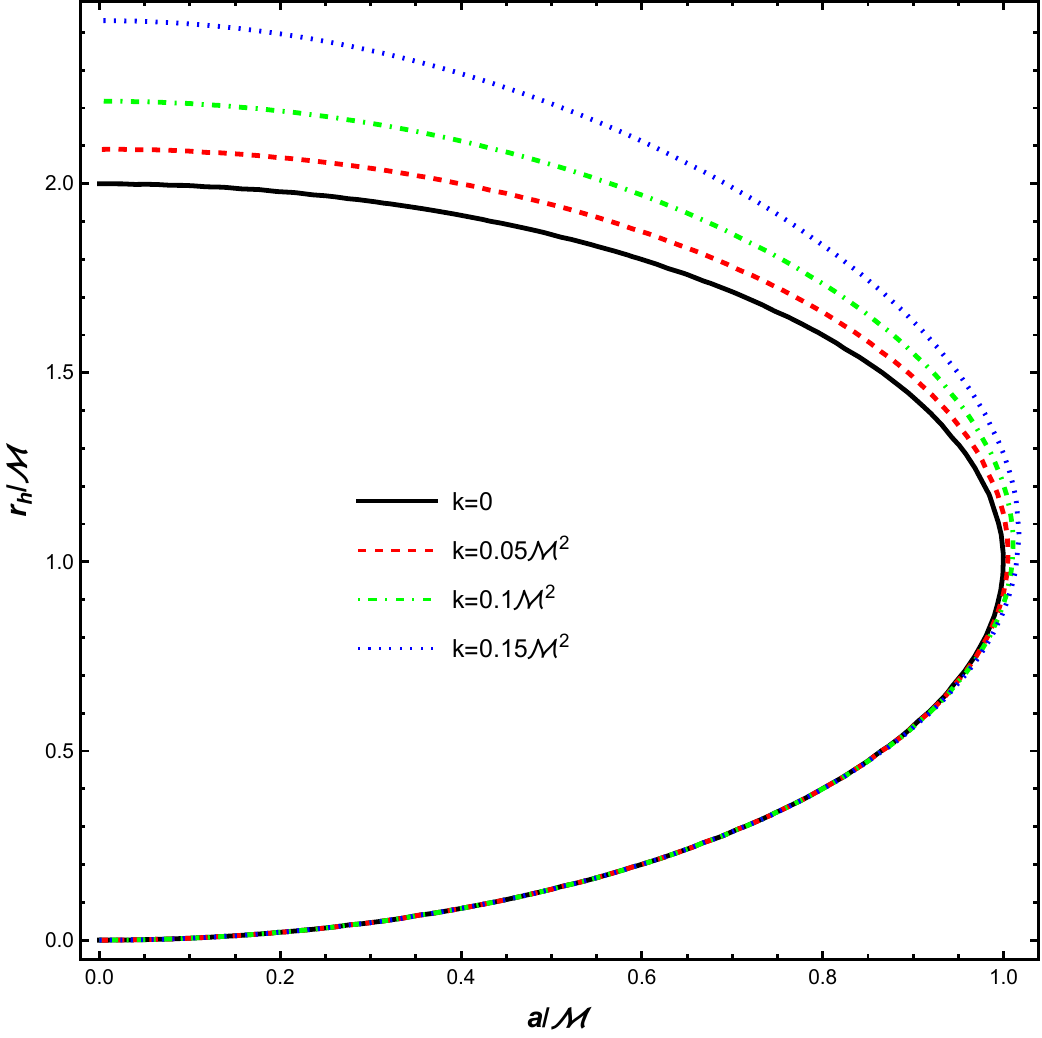}
    \includegraphics[scale=0.4]{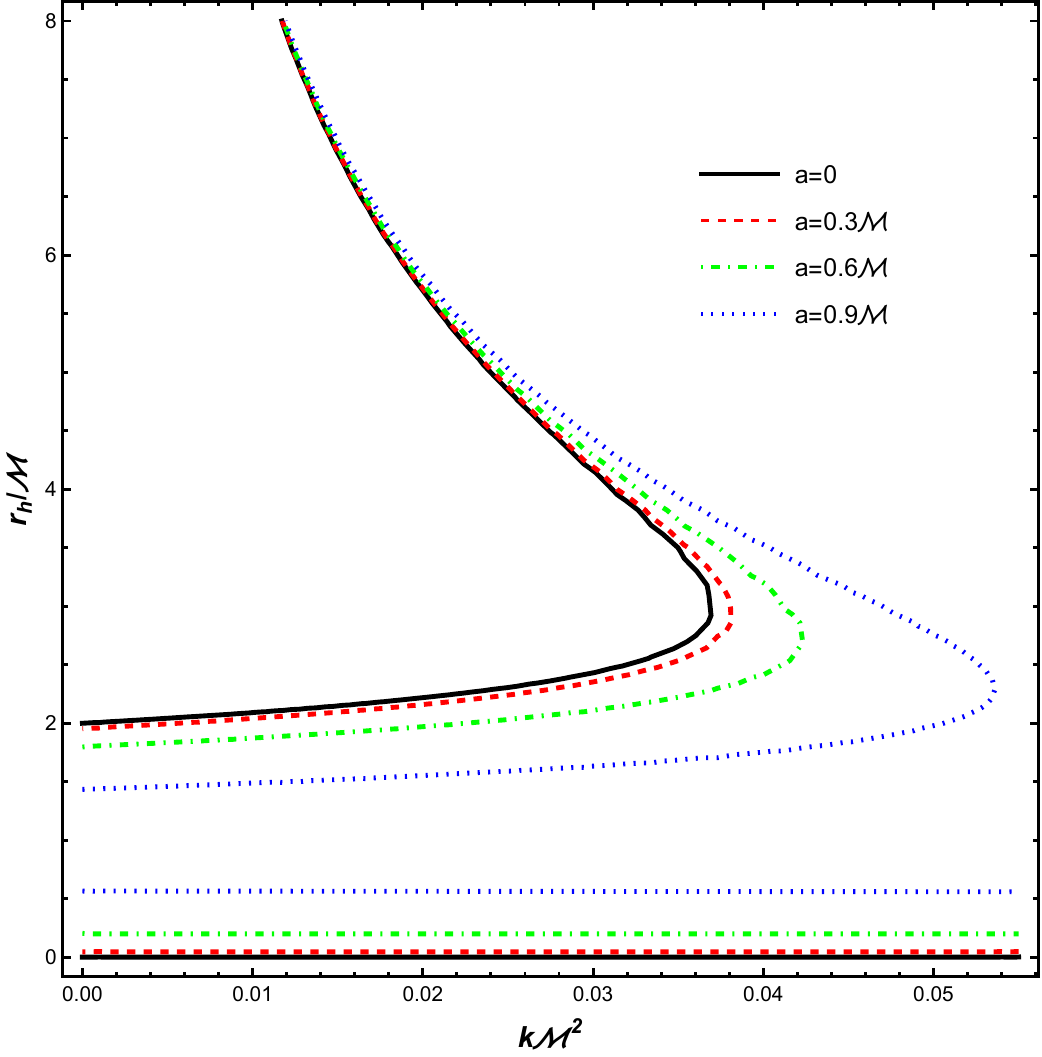}
    \caption{The depending of the event horizon to the BH spin parameter, with sufficiently small values of $\gamma$ and $\kappa$.}
    \label{fig:horizon structure}
\end{figure}
\ref{eq: outer horizon} and \ref{eq: inner horizon} show that the conformal Weyl correction shifts both the outer and inner horizons only perturbatively. The correction is governed by the dimensionless quantity $k M^2$, which is expected to be extremely small for astrophysical BHs. Consequently, the Kerr horizon structure is preserved, while the conformal parameters produce only a small displacement of the horizon locations. In limits $\gamma \rightarrow 0$ and $\kappa \rightarrow 0$, \ref{eq: outer horizon} and \ref{eq: inner horizon} reduce exactly to the Kerr horizons.
The extreme case is when event horizon (merges with inner horizon) is minimum and the null geodesics generating the horizon is zero, i.e.
\begin{equation}
    \nabla_r(n_\mu n^\mu)=0,
\end{equation}
this leads to the condition $\partial_r(\Delta^H(r))=0$ and approximate to $\gamma^2$ and $\kappa$ Weyl horizon and spin parameter are bia $M$:

\begin{align}
    r_+ &\rightarrow \mathcal{M}(1+2k\mathcal{M}^2) \approx M+2 \kappa  M^3-\frac{3}{2} \gamma  \left(M \left(6 \kappa  M^3+M\right)\right)+\frac{1}{2} \gamma ^2 \left(27 \kappa  M^5+M^3\right)+\mathcal{O}\left(\gamma ^3\right),\nonumber \;\; \\  a &\rightarrow \mathcal{M} \left(1+\frac{k \mathcal{M}^2}{2}\right) \approx M+\frac{\kappa  M^3}{2}-\frac{3}{4} \gamma  \left(M^2 \left(3 \kappa  M^2+2\right)\right)+\frac{1}{8} \gamma ^2 \left(27 \kappa  M^5+M^3\right)+\mathcal{O}\left(\gamma ^3\right). \label{eq:extreme case}
\end{align}

\eqref{eq:extreme case} demonstrates that the two conformal parameters play qualitatively different roles in determining the extreme BH configuration. The parameter $\kappa$ increases both the event horizon radius and the maximal allowed spin, indicating that the conformal correction permits slightly faster rotating BHs than in the Kerr geometry when $\kappa>0$. In contrast, a positive value of $\gamma$ shifts both quantities in the opposite direction, reducing the horizon radius and lowering the extreme spin. Consequently, $\gamma$ and $\kappa$ compete with each other in determining the extremality condition. In the limit $\gamma,\kappa \rightarrow 0$, the standard Kerr relation $r_+=a=M$ is exactly recovered.
In addition, we can also obtain the equation $r^2-3 \mathcal{M}+2a^2=0$ for the event horizon by reducing $k$ from the extreme conditions. Here, we choose a solution of $r$ that satisfies only the physical event horizon depended on $a$  as follows

\begin{equation}
    r_+=\frac{1}{2}\left(3\mathcal{M}+\sqrt{9 \mathcal{M}^2-8a^2}\right)\approx \frac{1}{2} \left(\sqrt{9 M^2-8 a^2}+3 M\right)-\frac{9}{4} \gamma  M^2 \left(\frac{3 M}{\sqrt{9 M^2-8 a^2}}+1\right)-\frac{81 \gamma ^2 a^2 M^4}{2 \left(9 M^2-8 a^2\right)^{3/2}} \label{extreme horizon to a}
\end{equation}

As can be seen from equation \eqref{extreme horizon to a}, in the extreme case, $a\rightarrow 3M(2-3M \gamma)/(4\sqrt{2})$ is allowed, in which case the event horizon accepts the smallest value $r_+=3M(2-3M \gamma)/4$ and creates the boundary condition $k \mathcal{M}^2=2/27$ for $k$.

We can also see from Fig.\ref{fig:horizon structure} (where the graphs are made by exact numerical solution of the equation $\Delta^\text{H}=0$) that the upper two panels are presented in terms of the original conformal parameters $\gamma$ and $\kappa$ in order to illustrate their independent influence on the horizon structure. Since the parameter $k$ is a specific combination of $\gamma$ and $\kappa$, expressing the horizons solely through $k$ would conceal the individual roles of the two conformal corrections. By contrast, the lower panels are plotted in terms of the effective mass $\mathcal{M}$ and the parameter $k$, which provide a more compact representation of the exact horizon equations and clearly demonstrate the global horizon structure of the Weyl spacetime. Despite of last plot the maximum values of $a$ define the extreme cases and the lines above it (decreasing) represent $r_+$ and the lines below it (increasing) represent $r_-$, which are added to the extreme case. The last panel shows the horizon structure bounded by $k$, we can see that the lower lines here represent the inner, above from it the outer (event) horizon, and at the top there is another positive horizon, which we can call it the de-Sitter horizon. The de-Sitter horizon is very far from the event horizon for sufficiently small values of $k$, we discuss processes near the event horizon in this paper, so we can ignore the de-Sitter horizon here. We can also see that the lower, left panel confirm the extreme case of \eqref{eq:extreme case}. Therefore, in such a case, we can assume that there are no root relations in the solution of the fourth-degree algebraic equation $\Delta^\text{H}=0$ and, if necessary, we can also use the following exact solutions when $k \neq 0$
\begin{eqnarray}
    r_+&=&\frac{\sqrt{A}}{2}+\frac{1}{2} \sqrt{-\frac{4\mathcal{M}}{\sqrt{A} k}-A+\frac{2}{k}} ,\;\;\text{outer event horizon}\label{outer horizon} \\ 
    r_-&=&\frac{\sqrt{A}}{2}-\frac{1}{2} \sqrt{-\frac{4\mathcal{M}}{\sqrt{A} k}-A+\frac{2}{k}},\;\;\text{inner horizon} \label{inner horizon}\\
    r_{dS}&=&-\frac{\sqrt{A}}{2}+\frac{1}{2} \sqrt{\frac{4\mathcal{M}}{\sqrt{A} k}-A+\frac{2}{k}},\;\;\text{de-Sitter horizon} \label{deSitter horizon}\\
    r_n&=&-\frac{\sqrt{A}}{2}-\frac{1}{2} \sqrt{\frac{4\mathcal{M}}{\sqrt{A} k}-A+\frac{2}{k}},\;\;\text{negative (non-physical)} \label{negative horizon}
\end{eqnarray}
where
\begin{eqnarray}
    A&=&-\frac{1-12 a^2 k}{3 D k}-\frac{D}{3 k}+\frac{2}{3 k},\nonumber \\ 
    D&=&\sqrt[3]{\sqrt{\left(36 a^2 k-54 k \mathcal{M}^2+1\right)^2-\left(1-12 a^2 k\right)^3}+36 a^2 k-54 k \mathcal{M}^2+1}.\nonumber
\end{eqnarray}
Solutions \eqref{outer horizon} and \eqref{inner horizon} are the outer and inner horizons around Kerr, while \eqref{deSitter horizon} is the de Sitter horizon (we discussed above that we do not work with this horizon and we can say a large distance that near point of this horizon), solution \eqref{negative horizon} takes negative real values even if it exists, and we do not take it as a physical horizon. %The extreme cases are $r_+=r_-=\sqrt{A}/2=\sqrt{1-\sqrt{1-12a^2k}}/\sqrt{6k}$, satisfying the condition $\left(36 a^2 k-54 k \mathcal{M}^2+1\right)^2=\left(1-12 a^2 k\right)^3$.

The static surface is found using the expression $g_{tt}=0$ and the region between the event horizon and the static surface is called the ergo-region. For a Weyl field, the ergo-region in the equatorial plane takes the form $r_+<r<2\mathcal{M}(1+4 k \mathcal{M}^2)$. 

\section{Equatorial equation of motion} \label{sec: Equatorial equation of motion}
In Weyl conformal geometry, the gauge-covariant derivative is compatible with the metric, $\nabla_\mu g_{\nu\rho} = 0$, which guarantees that parallel transport preserves the norm of any vector. As a consequence,
the invariant relation $P_\mu P^\mu = -m^2$ remains valid for a massive test particle, exactly as in Riemannian geometry. This ensures that the conservation of the particle’s norm and the associated energy-momentum law are respected in conformal gravity. In \cite{Condeescu:2025} explicitly shown that the conservation laws for the stress-energy tensor and the Weyl gauge current hold in a manifestly Weyl gauge covariant form, and are equivalent to the usual diffeomorphism invariance in the Riemannian picture.

It is known that conserved quantities are the first integrals of the equations motion a test particle. And, we can also introduce conserved quantities using Killing vectors $P_\mu\xi^\mu=\text{const}$. In axially symmetric fields, the time-like Killing vector $\xi^\mu_{(t)}=\partial_t=(1,0,0,0)$ and the space-like Killing vector $\xi^\mu_{(\phi)}=\partial_\phi=(0,0,0,1)$ represent the conserved energy $P_t=-E$ and the conserved angular momentum $P_\phi=L$, respectively. Here, $P_\mu=m\dot{x}^\mu$ is the 4-momentum, which expresses the invariant relation as follows
\begin{align}
    P_\mu P^\mu =-m^2, \;\; g_{rr}\dot{r}^2+g_{\theta\theta}\dot{\theta}^2=\frac{g_{\phi\phi}E^2+2g_{t\phi}EL+g_{tt}L^2}{m^2(g^2_{t\phi}-g_{tt}g_{\phi\phi})}-1=V(r,\theta), \label{eq:main motion}
\end{align}
where $m$ - is the rest mass of the test particle and dot means the derivative with respect to affine parameter. The case $\theta=\pi/2$ gives the equation of radial motion in the equatorial plane $\dot{r}^2=V(r,\pi/2)/g_{rr}=V_\text{eff}(r)$, where $V_\text{eff}$ is called the radial effective potential. The $V_\text{eff}$ is classically equivalent to the potential energy and $V_\text{eff}=0$ denotes the turning points of the test particle and in addition to this condition $\partial_r V_\text{eff}=0$ the circular orbits are defined. These orbits are considered stable when $\partial^2_r V_\text{eff}<0$, and inner stable circular orbits (ISCO) is found from the condition $\partial^2_r V_\text{eff}=0$.
\begin{figure}
    \centering
    \includegraphics[scale=0.4]{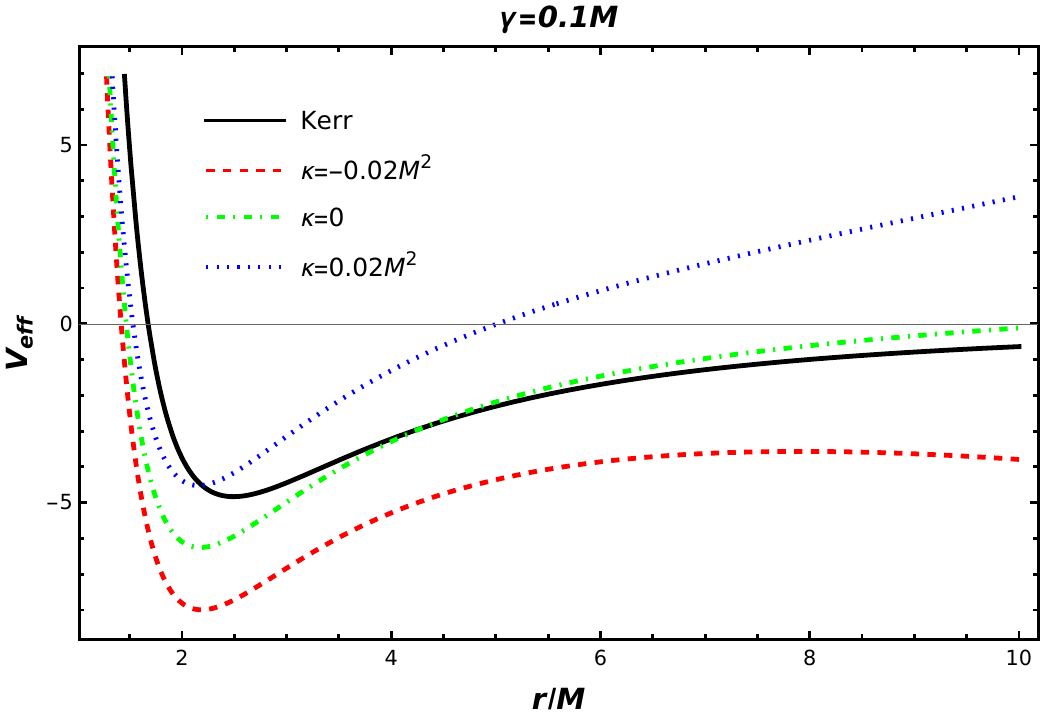}
    \includegraphics[scale=0.4]{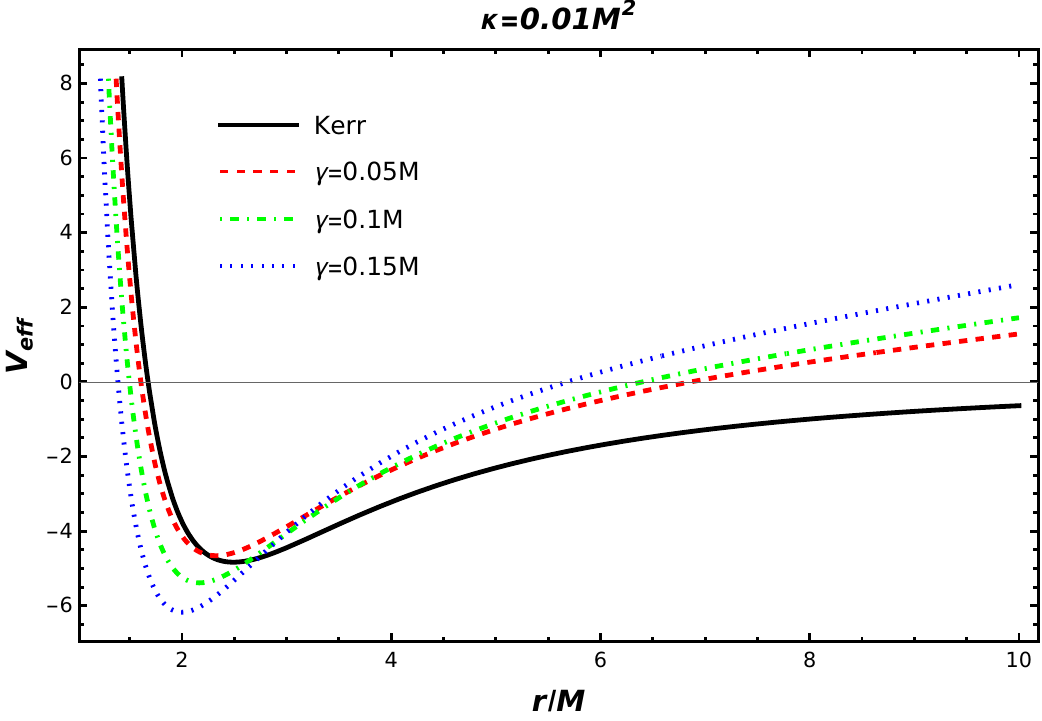}
    \includegraphics[scale=0.4]{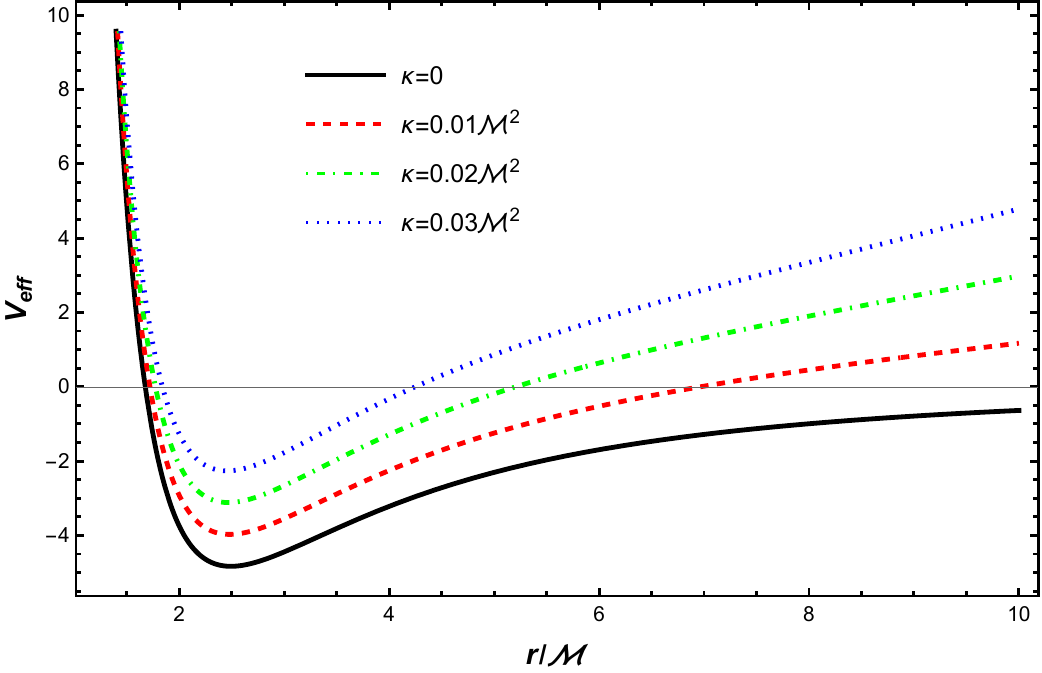}
    \includegraphics[scale=0.4]{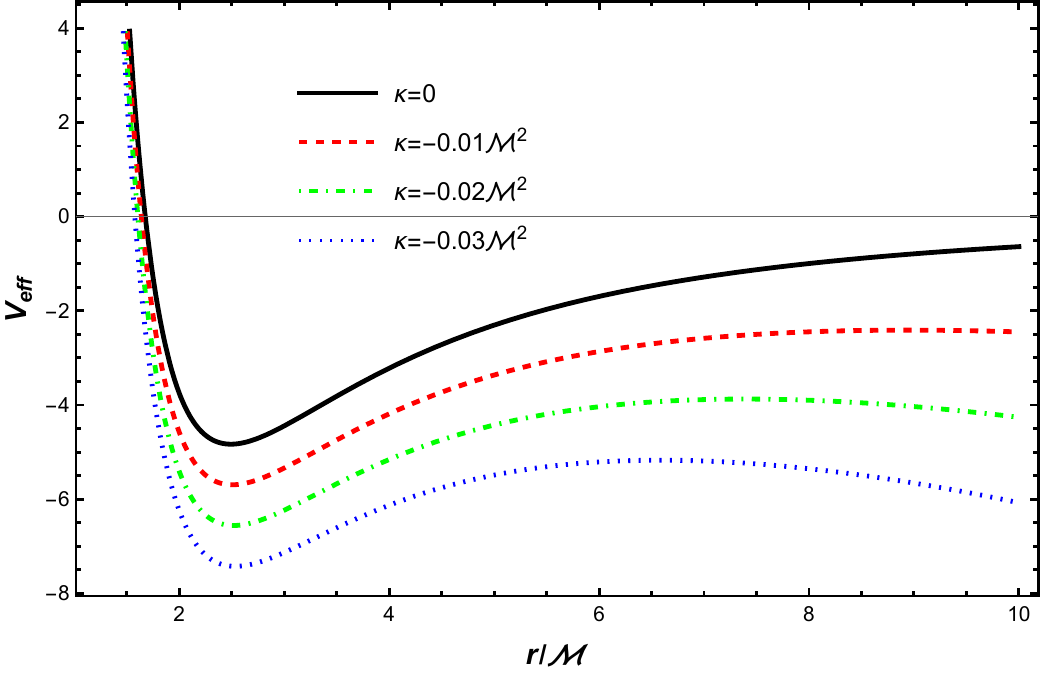}
    \caption{The relationship between the effective potential and the radial coordinate in two frames: $V_\text{eff}=V_\text{eff}(r,M,\gamma,\kappa)$ and $V_\text{eff}=V_\text{eff}(r,\mathcal{M},k)$. Here we choose $L/m=10M (\mathcal{M})$. }
    \label{fig:effective potential}
\end{figure}

Fig.\ref{fig:effective potential} illustrates the radial dependence of the effective potential $V_{\rm eff}$ for a massive test particle ($E = m$) in the equatorial plane and in the extreme Weyl rotating BH, with fixed angular momentum $L/m = 10M (\mathcal{M)}$. The upper left panel (upper panels show the $V_\text{eff}=V_\text{eff}(r,M,\gamma,\kappa)$ frame) shows the effect of the cosmological parameter $\kappa$ at fixed $\gamma = 0.1M$, while the upper right panel demonstrates the influence of $\gamma$ at fixed $\kappa = 0.01M^2$. In both panels, the standard Kerr result (solid black curve) is shown for reference. The local minima and maxima of $V_{\rm eff}$ correspond to stable and unstable circular orbits, respectively, while the zeros of the effective potential indicate the turning points of the radial motion. It is evident that increasing $\kappa$ shifts the inner turning point outward, whereas $\gamma$ primarily affects the depth of the potential well and the location of the innermost stable circular orbit (ISCO). The bottom two panels show the $V_\text{eff}=V_\text{eff}(r,\mathcal{M},k)$ frame, where the lines do not intersect at different values of $k$, unlike the top panels. The bottom left panel represents the cases $k>0$, where as $k$ increases,  the inner turning points move outwards and the outer turning points move inwards. In the cases $k<0$, the inner turning points move inwards by a small amount, and we can see that there are no outer turning points in this case. Additionally, In the upper panels, one observes that for $\kappa>0$ both the inner and outer turning points shift inward. This displacement can be interpreted as the effect of a background cosmological force acting on the test particle: the positive $\kappa$ parameter effectively introduces a de-Sitter–like repulsion that modifies the radial potential well. In contrast, the lower panels expressed in terms of $(\mathcal{M}, k)$ emphasize the cumulative impact of this background correction, showing that the turning points move inward in a way that is absent in the standard Kerr geometry. These shifts therefore represent genuine signatures of Weyl conformal gravity, where the cosmological background directly influences particle dynamics near the BH.
\section{Extractable energy and PP} \label{sec: Extractable energy and PP}
Energy extraction processes around rotating BHs represent one of the most important consequences of relativistic frame dragging in strong gravitational fields. Among these mechanisms, the original PP provides a classical way to extract rotational energy from a BH through particle interactions inside the ergoregion. The existence of negative-energy states in this region allows part of the BH rotational energy to be transferred to escaping particles, leading to a decrease in the BH angular momentum. The efficiency of this process strongly depends on the spacetime geometry, the structure of the event horizon, and the properties of the ergosphere. Therefore, modified theories of gravity may significantly affect the energy extraction mechanism through corrections to the Kerr geometry. In conformal Weyl gravity, additional conformal parameters modify the rotating BH spacetime and can influence the extractable energy, particle trajectories, and the efficiency of the PP. In this section, we investigate the extractable energy and the original Penrose scenario around rotating Weyl BHs.
\subsection{Extractable energy} \label{subsec: Extractable energy}
The PP is a classical method for extracting energy from rotating BHs, and so we will first discuss the energy that can be extracted from Weyl BHs. The energy that can be extracted from rotating BHs is found by subtracting the irreducible energy $M_\text{irr}$ from the total BH mass-energy $M$ . The irreducible energy is the total energy inside the event horizon of the BH, which is defined by the Christodoulou-Ruffini-Hawking \cite{PhysRevD.4.3552, PhysRevLett.25.1596} formula
\begin{eqnarray} 
    M^2_\text{irr}&=&\frac{1}{16\pi}\int\limits_0^{2\pi}d\phi\int\limits_0^{\pi}\sqrt{g_{\theta\theta}g_{\phi\phi}}d\theta=\frac{1}{4}\left(r^2_++a^2\right)\approx 
    \frac{1}{2} \left(\mathcal{M} \sqrt{\mathcal{M}^2-a^2}+\mathcal{M}^2\right)+\frac{k \left(\sqrt{\mathcal{M}^2-a^2}+\mathcal{M}\right)^5}{4 \sqrt{\mathcal{M}^2-a^2}}.\label{irreducible mass}
\end{eqnarray}
To calculate the total mass-energy of a conformal Weyl BH, we use the Komar mass \cite{1963PhRv..129.1873K} expression
\begin{align}
    M_\text{Komar}=-\frac{1}{8\pi} \oint\limits_{S_\infty}\nabla^\mu\xi^\nu dS_{\mu\nu}=-\frac{1}{8\pi}\int\limits^{2\pi}_0 d\phi \int\limits^\pi_0 (\partial_rg_{tt}) r^2 \sin\theta d\theta=\mathcal{M} \label{Komar mass}
\end{align}
where $\xi^\mu=\partial_t=(1,0,0,0)$ is the time-like Killing vector. In \eqref{Komar mass} the integration is taken for a surface at infinity. The last term of $g_{tt}$ in \eqref{rotating metric} gives the cosmological background energy flux at infinity, and on an infinitely large scale this energy is divergent, so the contributions of these terms can be discarded. In a similar way, we introduce the $\eta^\mu=\partial_\phi=(0,0,0,1)$ -space-like Killing vector and it is easy to see that the Komar angular momentum is 
\begin{equation}
L_\text{Komar}=-\frac{1}{8\pi}\oint\limits_{S_{\infty}}\nabla^\mu\eta^\nu dS_{\mu\nu}=\frac{1}{16\pi}\int\limits^{2\pi}_0 d\phi \int\limits^\pi_0 g^{rr}\left(g^{tt}\partial_rg_{t\phi}+g^{t\phi}\partial_r g_{\phi\phi}\right) r^2 \sin\theta d\theta=M_\text{Komar} a=\mathcal{M} a. 
\end{equation}
Then, in Weyl gravity the extractable energy we can found by $M_\text{ext}=M_\text{Komar}-M_\text{irr}$ as 
\begin{align}
    M_\text{ext}=\mathcal{M}-\sqrt{\frac{\mathcal{M}\left(\mathcal{M}+\sqrt{\mathcal{M}^2-a^2}\right)}{2}}\left(1+\frac{k \left(\mathcal{M}+\sqrt{\mathcal{M}^2-a^2}\right)^4}{4\mathcal{M}\sqrt{\mathcal{M}^2-a^2}}\right). \label{eq:extractable energy}
\end{align}
The \eqref{eq:extractable energy} shows the energy that can be extracted from the non-extreme Weyl BH. This energy is the largest in the extreme case, using \eqref{eq:extreme case} we recalculate \eqref{irreducible mass} and express the extractable energy for the extreme case as
\begin{align}
    M_\text{ext}=\mathcal{M}-\frac{\mathcal{M}}{\sqrt{2}}-\frac{5k\mathcal{M}^3}{4\sqrt{2}}\approx \frac{1}{2} \left(2-\sqrt{2}\right) M -\frac{3}{4} \left(2-\sqrt{2}\right) \gamma  M^2 -\frac{5 M^3 \left(\gamma ^2+4 \kappa \right)}{16 \sqrt{2}} \label{extractable energy in extreme cas}
\end{align}

The first term definite value in \eqref{extractable energy in extreme cas} represents the largest extractable energy in the Kerr geometry, which is approximately $0.293$ of the total BH energy, while the other terms is an additional term for conformal Weyl gravity, which reduces the extractable energy. 
\subsection{Original Penrose scenario} \label{subsec: Original Penrose scenario}
Let's consider the original PP around a conformal Weyl BH. A particle coming (incoming particle) from large distance in the direction of the BH's rotation (direct orbit) splits into two at the point of closest turning point to the event horizon, one of them (the first particle) moves in the opposite direction to the direction of the BH's rotation (retrograde orbit) and, being in the ergo-region, has negative energy and falls into the BH, cutting across the event horizon. The other particle (the second particle) moves in the direction of the BH's rotation and escapes the BH's influence with a greater energy than the incoming particle \cite{1969NCimR...1..252P}. In the ideal case, the resulting particles are considered ultrarelativistic $(m=0)$. To mathematically consider these considerations, we can determine the angular momentum of particles using by \eqref{eq:main motion} in the case $\dot{\theta}=0$ 
\begin{equation}
    L=-\frac{E g_{t\phi}+\sqrt{(g^2_{t\phi}-g_{tt}g_{\phi\phi})(E^2+m g_{tt})}}{g_{tt}}, \;\; L_1=E_1\frac{g_{t\phi}-\sqrt{g^2_{t\phi}-g_{tt}g_{\phi\phi}}}{g_{tt}}, \;\; L_2=-E_2\frac{g_{t\phi}+\sqrt{g^2_{t\phi}-g_{tt}g_{\phi\phi}}}{g_{tt}} \label{E and L connect in Penrose}
\end{equation}
and energies of the 1 and 2 particles are found from the conservation of the angular momentum
\begin{align}
    E_1=\frac{1}{2}\left(E+\sqrt{E^2+m^2g_{tt}}\right), \;\;\; E_2=\frac{1}{2}\left(E-\sqrt{E^2+m^2g_{tt}}\right) \label{energies of 1 and 2 particles}
\end{align}
In this process, we assume that the turning point where the particle comes closest to the BH is at the event horizon. Finally, we easily obtain the energy efficiency by considering $\Delta^\text{H}=0$ as
\begin{equation}
    \epsilon=\frac{E_1-E}{E}=-\frac{E_2}{E}=\frac{1}{2}\left(\sqrt{1+\frac{m^2 g_{tt}}{E^2}}-1\right)=\frac{1}{2}\left(\sqrt{1+\frac{m^2}{E^2}\frac{a^2 \sin^2{\theta}-a^4 k \cos^4{\theta}}{r_+^2+a^2 \cos^2{\theta}}}-1\right). \label{energy efficiency in Penrose}
\end{equation}
If we take into account that the constants $\gamma$ and $\kappa$ are very small here, it can be seen from \eqref{energy efficiency in Penrose} that the $a$-spin parameter dominates for the largest energy efficiency, and this value is largest only at the equator ($\theta=\pi/2$) in the $\dot{\theta}=0$ planes. The value of $a$ is largest in extreme cases, and in this case the horizon is smallest. The smallest energy of the incoming particle energy at infinity is $E=m$ at the background. Based on these considerations, we rewrite \eqref{energy efficiency in Penrose} as
\begin{eqnarray}
    \epsilon= \frac{1}{2} \left(\sqrt{1+\frac{1+k\mathcal{M}^2}{1+4k\mathcal{M}^2}}-1\right)\approx \frac{1}{\sqrt{2}}-\frac{1}{2}-\frac{3k\mathcal{M}^2}{4 \sqrt{2}}\approx \frac{1}{2} \left(\sqrt{2}-1\right)-\frac{3 M^2 \left(\gamma ^2+4 \kappa \right)}{16 \sqrt{2}} \label{evaluated efficiency in penrose}
\end{eqnarray}
In equation \eqref{extractable energy in extreme cas}, we ca see that in the extreme cases, the extractable energy to be less due to $\gamma$, while in the PP, we can see that it can be reduced only by $k$. Moreover, as can be seen from the last term of \ref{evaluated efficiency in penrose}, the reduction is small, approximately on the order of $\kappa$ or $\gamma^2$, and if the $k<0$ case dominates, the energy efficiency can be partially increased. If we consider \eqref{extreme horizon to a}, we have exactly efficiency $\epsilon= (\sqrt{6}-2)/4\approx 0.112$ when $k\mathcal{M}^2=2/27$.
\subsection{BH slowdown}\label{BH slow down}
In the PP, the second particle crossing the event horizon gives the BH an additional angular momentum $dL=L_2$ and an additional energy $d\mathcal{M}=E_2$. In this case, the BH can be considered to have done work in the external process, and as a result, its rotation slows down. Using \eqref{E and L connect in Penrose}, the equation of the slowing down process can be written as
\begin{equation}
    \frac{d(a\mathcal{M})}{d\mathcal{M}}=\frac{L_1}{E_1}=-\frac{g_{t\phi}}{g_{tt}}=\frac{r^2_++a^2}{a} \label{slowdown equation}
\end{equation}
here, we considered $g^2_{t\phi}-g_{tt}g_{\phi\phi}=0$ at the event horizon. It is known that as a result of any external influence, in the BH must be $\delta M_{irr} \ge 0$. Since we are considering the PP for ideal cases, we can take $\delta M_{irr}=0$ here, and in the Kerr geometry this condition is exactly fulfills. Therefore, based on \eqref{irreducible mass}, the solution of \eqref{slowdown equation} can be obtained in the form $r^2_++a^2=\text{const}$. The slowdown law for the Kerr geometry is the form $a=M_0\sqrt{2-M^2_0/M^2}$, where $M_0$ is the mass of the BH in the extreme case. To find this law for the Weyl BH, we consider $r^2_++a^2=\mathcal{M}^2_0\left(2+5k\mathcal{M}_0^2\right)$ in the extreme case.
\begin{equation}
\frac{a^2}{\mathcal{M}^2_0}=2+5k\mathcal{M}^2_0-\frac{\mathcal{M}^2_0}{\mathcal{M}^2}\left(1+4k \mathcal{M}^2_0\right) \label{slowdown in Weyl}
\end{equation}
For the Kerr BH to stop rotating, its mass must decrease to $M=M_0/\sqrt{2}$, while in Weyl gravity this value is equal to $\mathcal{M}\approx \mathcal{M}_0(1+3k\mathcal{M}^2_0/4)/\sqrt{2}$. So, if $k>0$, the Weyl BH can lose less energy when it stops rotating, otherwise it can lose more energy.

In the PP, the energy of 2 particles with negative energy crossing the event horizon is added to the BH $dM=E_2$ and we get the mass of the incoming particle  as $dm=E$. Now, using second equation of \eqref{energies of 1 and 2 particles} and \eqref{slowdown in Weyl}, we determine the total mass of the massive particles involved in the PP, when the Weyl BH mass decreases from $\mathcal{M}_0$ to $\mathcal{M}$.
\begin{eqnarray}
    \frac{m}{\sqrt{2}\mathcal{M}_0}=\int\limits^\mathcal{M}_{\mathcal{M}_0}\frac{\sqrt{2}d(\mathcal{M}/\mathcal{M}_0)}{1-\sqrt{1+\frac{a^2}{r^2_+}}}=\log{\frac{\mathcal{M}_0(\sqrt{2}-1)}{\sqrt{2}\mathcal{M}-\mathcal{M}_0}}-\frac{k \mathcal{M}^2_0}{4}\left(9\sqrt{2}+9-\frac{9\mathcal{M}_0}{\sqrt{2}\mathcal{M}-\mathcal{M}_0}-\log{\frac{M_0(\sqrt{2}-1)}{\sqrt{2}M-M_0}}-4\log{\frac{\mathcal{M}}{\mathcal{M}_0}}\right) \label{eq: mass of massive particles}
\end{eqnarray}
Expression \eqref{eq: mass of massive particles} is approximated to the first order of $k$, and the first term in the resulting expression gives the Kerr BH result at $\gamma=0$. In the Kerr model, the mass of particles required for a BH to stop rotating is infinite. For the Weyl BH, the occurrence of the terms $m \sim - \log{k}$ in the approximation, when it stops rotating, also means that in this model, an infinite number of particles are required for the PP.

\section{BH Thermodynamics} \label{sec: BH thermodynamics}
BH thermodynamics provides one of the deepest connections between gravitation, quantum theory, and statistical physics, indicating that BHs behave as genuine thermodynamic systems characterized by temperature, entropy, and conserved thermodynamic quantities. In particular, the existence of Hawking radiation implies that BHs are not completely classical objects, but instead possess microscopic degrees of freedom whose nature is expected to be explained by a fundamental quantum theory of gravity. The thermodynamic behavior of BHs also plays an essential role in understanding spacetime stability, phase transitions, and the interplay between geometry and quantum effects in strong gravitational fields. 

In higher-curvature and conformally invariant theories of gravity, the study of BH thermodynamics becomes especially important because additional geometric terms in the gravitational action generally modify the standard relations between entropy, temperature, and conserved charges. As a consequence, the thermodynamic structure of BHs in such theories can substantially differ from the corresponding Einstein gravity results. In conformal Weyl gravity, the presence of higher-derivative curvature contributions and conformal symmetry may alter the horizon structure, surface gravity, stability conditions, and critical behavior of rotating BHs. Therefore, investigating the thermodynamic properties of Weyl BHs provides important insight into the physical consistency of conformal gravity and the influence of conformal corrections on BH mechanics. In this section, we analyze the thermodynamic characteristics of rotating Weyl BHs and examine the validity of the four laws of BH thermodynamics within the framework of conformal Weyl gravity.
\subsection{Zeroth law}
According to the zeroth law of thermodynamics, the surface gravity - $\kappa_g$ above the event horizon must be constant. To determine the surface gravity, we use the $K^\mu \partial_\mu=\partial_t+\Omega_H\partial_\phi$ -Killing vectors, where $\Omega_H=-g_{t\phi}/g_{\phi\phi}=a/(r^2_++a^2)$ is the angular velocity of the event horizon and square of this Killing vektors is $K^2=-K^\mu K_\mu=0$ at the event horizon. Then, we can take the surface gravity from 
\begin{equation}
K^\mu \nabla_\mu K^\nu=\kappa_g K^\nu. \label{surface gr eq} 
\end{equation}
We simplify equation \eqref{surface gr eq} using the following equations 
\begin{equation}
    \nabla_\mu K_\nu=-\nabla_\nu K_\mu \,,\;\; \nabla_\mu K^2=2 K^\nu\nabla_\nu K_\mu=2\kappa_g K_\mu \label{Killing equations}
\end{equation}
that are valid for Killing vectors. We take the square of the second equation in \eqref{Killing equations} and the following equation for surface gravity is obtained
\begin{equation}
    \kappa^2_g=-\frac{ g^{\mu\nu}(\partial_\mu K^2)(\partial_\nu K^2)}{4 K^2} \Bigg|_{r=r_+}
\end{equation}
for the Weyl gravity
\begin{equation}
    -K^2=\left[ (r^2+a^2)\Omega _H-a\right]^2 \frac{\Delta^\theta \sin^2 \theta}{\rho^2}-\left[a\Omega_H \sin^2\theta-1\right]^2 \frac{\Delta^H}{\rho^2}. \label{Killing square}
\end{equation}
It is clear from \eqref{Killing square} that only the derivative with respect to $r$ is nonzero above the event horizon, so
\begin{eqnarray}
    \kappa_g&=&\frac{1}{2} \sqrt{-\frac{g^{rr}  \left( \partial_r K^2\right)^2}{K^2}}\Bigg|_{r=r_+}=\frac{1}{2} \frac{\partial_r \Delta^H}{r^2_++a^2}\Bigg|_{r=r_+} \approx\nonumber \\
      &&\frac{\sqrt{\mathcal{M}^2-a^2}}{2\mathcal{M}\left(\mathcal{M}+\sqrt{\mathcal{M}^2-a^2}\right)}
    -\frac{ \left(\sqrt{\mathcal{M}^2-a^2}+\mathcal{M}\right)^2 \left(4 \mathcal{M} \sqrt{\mathcal{M}^2-a^2}-a^2\right)}{4\mathcal{M}^2 \sqrt{\mathcal{M}^2-a^2}}k\approx \nonumber\\ 
    &&\frac{\sqrt{M^2-a^2}}{2M\left(M+\sqrt{M^2-a^2}\right)} +\frac{3 \left(a^4+M^3 \sqrt{M^2-a^2}-M^4\right)}{4 a^2 \left(a^2-M^2\right)}\gamma+\frac{\left(a^2-4 M \sqrt{M^2-a^2}\right) \left(\sqrt{M^2-a^2}+M\right)^2}{4 M^2 \sqrt{M^2-a^2}}\kappa  \nonumber \\ 
    &&+\frac{M^6 \left(10 M-17 \sqrt{M^2-a^2}\right)+2 a^2 M^4 \left(9 \sqrt{M^2-a^2}-7 M\right)+a^6 \left(\sqrt{M^2-a^2}+6 M\right)}{16 \left(M^3-a^2 M\right)^2}\gamma^2 \nonumber \\
    &&-\frac{a^4 M^2 \left(11 \sqrt{M^2-a^2}+2 M\right)}{16 \left(M^3-a^2 M\right)^2}\gamma^2+
    \mathcal{O}(\gamma^3)+\mathcal{O}(\kappa^2)
    \label{surface gravity}
\end{eqnarray}
From \eqref{surface gravity} we can see that a rotating Weyl BH satisfies the zeroth law of thermodynamics. Using \eqref{eq:extreme case}, it is easy to see that in extreme cases $\kappa_g=0$. Also, \eqref{surface gravity} shows explicitly how the conformal parameters modify the surface gravity with respect to the Kerr result. The first term reproduces the Kerr expression in the limit $\gamma,\kappa \rightarrow 0$, while the subsequent terms represent the corrections induced by the conformal parameters. In particular, $\gamma$ and $\kappa$ contribute independently to the surface gravity already at first order, whereas higher-order terms in $\gamma$ provide additional corrections near the extremal regime. These corrections modify both the magnitude of the surface gravity at a fixed spin.
\begin{figure}
    \centering
    \includegraphics[scale=0.5]{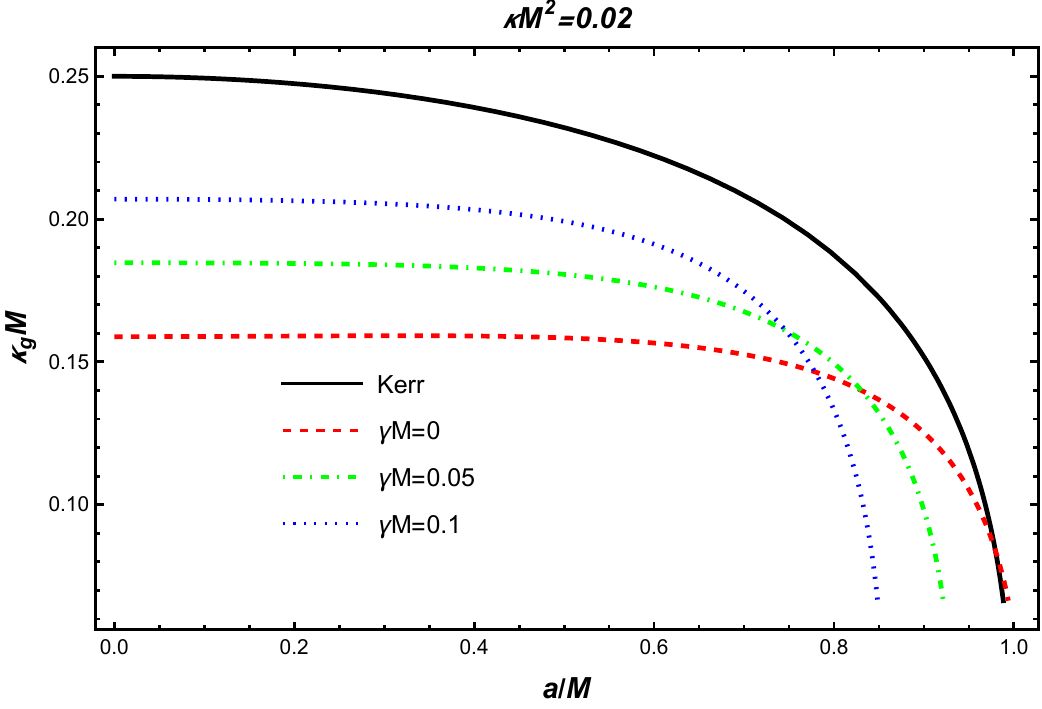}
    \includegraphics[scale=0.5]{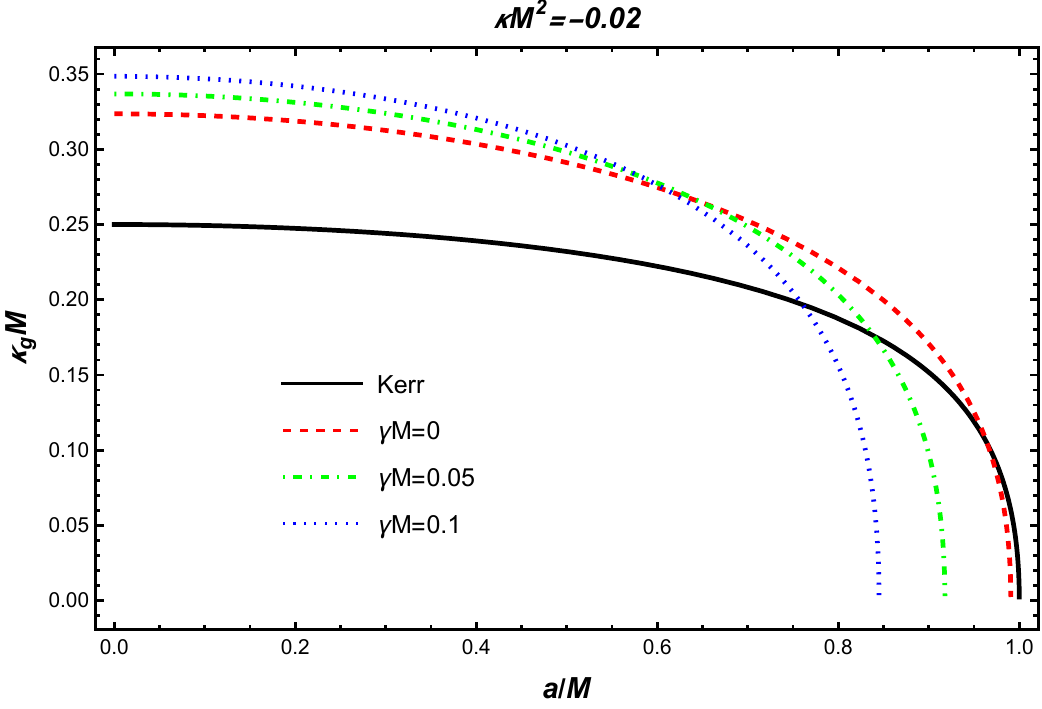}
    \caption{Dimensionless surface gravity $\kappa_g M$ as a function of the dimensionless spin parameter $a/M$ for different values of the conformal parameter $\gamma$. The left panel corresponds to $\kappa M^2=0.02$, while the right panel corresponds to $\kappa M^2=-0.02$. The solid black curve represents the Kerr case, whereas the dashed, dash-dotted, and dotted curves correspond to $\gamma M=0, \, 0.05$ and $0.1$ respectively. The vanishing of $\kappa_g$ identifies the extreme configuration. The comparison demonstrates that the effect of $\gamma$ on the surface gravity depends on the sign of $\kappa$.}
    \label{fig:surface gravity}
\end{figure}

\subsection{First law} \label{subsec: first law}
Before discussing the first law of thermodynamics, we need to find the Weyl BH entropy. The BH entropy is a quantity that combines the gravitational and quantum effects of a BH and characterizes its internal degrees of freedom. Since the Wald BH is conformal invariant, we cannot calculate the entropy directly using the $S=A/4$,  Bekenstein-Hawking formula \cite{1973PhRvD...7.2333B, 1975CMaPh..43..199H}, where, $A=4\pi(r^2_++a^2)$ is the event horizon $2D$ surface. We calculate the entropy using the following Wald formula \cite{Wald_2001}
\begin{eqnarray}
    S=-2\pi\int\limits_{A}\frac{\partial \mathcal{L}}{\partial R_{\mu\nu\sigma\lambda}}\varepsilon_{\mu\nu}\varepsilon_{\sigma\lambda} dA \label{Wald entropy}
\end{eqnarray}
 here $\varepsilon_{\mu\nu}=\nabla_\mu K_\nu/\kappa_g$  is the bi-normal tensor to the event horizon surface, and normalization condition is $\varepsilon_{\mu\nu}\varepsilon^{\mu\nu}=-2$. For conformal Weyl gravity the Lograngian density as follows 
\begin{equation}
\mathcal{L}_W=\alpha_{\mathrm{g}}\left(R^{\mu\nu\sigma\lambda}R_{\mu\nu\sigma\lambda}-2R^{\mu\nu}R_{\mu\nu}+\frac{1}{3}R^2\right)\label{Lograngian density}
\end{equation}
The metric in the form \eqref{rotating metric} also satisfies the Einstein equation in the form
\begin{eqnarray}
    R_{\mu\nu}-\frac{1}{2}Rg_{\mu\nu}+3k g_{\mu\nu}=0 . \label{Eynshteyn equation}
\end{eqnarray}
By contracting equation \eqref{Eynshteyn equation}, we define the relations $R=12k$ and $R_{\mu\nu}=3k g_{\mu\nu}$, and it is not difficult to see that these relations satisfy the equality $\mathcal{W}_{\mu\nu}=0$. 
So, considering \eqref{Wald entropy}, \eqref{Lograngian density}  and \eqref{Weyl tensor} we take the Weyl entropy expression in conformal gravity as
\begin{eqnarray}
    S_W&=&-4\pi \alpha_{\mathrm{g}}\int\limits_{A}\left(R^{\mu\nu\rho\sigma}- \frac{1}{2}\left(g^{\mu\rho} R^{\nu\sigma}-g^{\mu\sigma} R^{\nu\rho}+g^{\nu\sigma} R^{\mu\rho}-g^{\nu\rho} R^{\mu\sigma}\right)+\frac{1}{6} R\left(g^{\mu\rho} g^{\nu\sigma}-g^{\mu\sigma} g^{\nu\rho}\right) \right) \varepsilon_{\mu\nu} \varepsilon_{\rho\sigma} dA\nonumber\\ 
    &=&32\pi^2 \alpha_{\mathrm{g}} \int\limits_0^\pi\left(\frac{4 r_+^2 (a^2 + r_+^2 - k r_+^4)}{\left(r_+^2+a^2\cos^2{\theta}\right)^3}-\frac{a^2 + 4 m r_+ + r_+^2 - k r_+^4}{\left(r_+^2+a^2\cos^2{\theta}\right)^2}\right) (r^2_++a^2)\sin{\theta}d\theta=\frac{128\pi^2\alpha_g \mathcal{M}r_+}{r^2_++a^2}
\end{eqnarray}

The BH temperature is the radiation produced by quantum fluctuations or photon tunneling near the event horizon, and is intrinsically linked to the BH geometry. Spacetime is considered flat (Minkowski) near the horizon. When converted to the Euclidean metric, the time coordinate is periodized with a period $\beta=2\pi/\kappa_g$ to eliminate the "conical singularity" at the horizon, where $T=1/\beta$, and the Hawking temperature is characterized by $T=\kappa_g /(2\pi)$ surface gravity. Fig.\ref{fig:surface gravity} shows the dimensionless surface gravity $\kappa_g M$ as a function of the dimensionless spin $a/M$ for different values of the conformal parameter $\gamma$ according to \eqref{outer horizon}. The two panels correspond to positive and negative values of $\kappa$, $\kappa M^2=0.02$ and $\kappa M^2=-0.02$, respectively. For $\kappa M^2=0.02$, increasing $\gamma$ decreases the surface gravity at a fixed value of $a/M$, whereas for $\kappa M^2=-0.02$, the same increase in $\kappa$ enhances the surface gravity. Thus, the effect of the conformal parameter $\gamma$ is sensitive to the sign of $\kappa$, demonstrating a nontrivial interplay between the two conformal parameters. In both cases, the zero of the surface gravity identifies the extremal configuration, and its shift with $\gamma$ reflects the corresponding modification of the extremality condition.

If conformal Weyl model also satisfies \eqref{Eynshteyn equation}, the total action can be considered as Einstein-Hilbert+Weyl corresponding on EFT. Then, the totally conserved energy $M_\text{total}=M_\text{Komar}+M_\text{W}$, angular momentum $J_\text{total}=J_\text{Komar}+J_\text{W}$ and the entropy $S_\text{total}=A/4+S_\text{W}$ return to their usual values in the Kerr limit. 
According to the Wald formalism \cite{PhysRevD.48.R3427}, for any diffeomorphism-invariant theory of gravity, the first law of BH thermodynamics, $dM_\text{total}=TdS_\text{total}+\Omega_\mathrm{H} dJ_\text{total}$, satisfies by construction (depending on the choice of $\alpha$) and this also implies that the Weyl formalism cannot be an independent thermodynamic system. In the context of Weyl gravity, the inclusion of higher-derivative terms necessitates a consistent redefinition of both the entropy and the gravitational mass to preserve this thermodynamic identity. Here, $M_\mathrm{W}$ and $J_\mathrm{W}$ is the Wald mass and Wald angular momentum respectively, we can take them with Noether current in the conformal Weyl gravity as \cite{Peng_2014}
\begin{eqnarray}
    Q_\mathrm{W}=\frac{1}{8\pi}\int\limits^1_0\int\limits_\Sigma \left(\frac{1}{2}\delta K^{\mu\nu}+\frac{1}{4} K^{\mu\nu}g^{\alpha\beta}\delta g_{\alpha\beta}-\xi^{[\mu}\Theta^{\nu]}\right)d\Sigma_{\mu\nu}, \label{Wald mass int}
\end{eqnarray}
where
\begin{eqnarray}
    K^{\mu\nu}&=&4\alpha C^{\mu\nu\rho\sigma}\nabla _{[\rho}\xi_{\sigma]}-8\alpha \xi_\sigma \left(\nabla^{[\mu}R^{\nu]\sigma}+\frac{1}{6} g^{\sigma[\mu}\nabla^{\nu]}R\right) \label{K tensor},\\
     \Theta^{\mu}&=&4\alpha C^{\mu\nu\rho\sigma}\nabla _{\sigma}\delta g_{\nu\rho}+4\alpha \delta g_{\nu\sigma} \left(\nabla^{[\mu}R^{\nu]\sigma}+\frac{1}{6} g^{\sigma[\mu}\nabla^{\nu]}R\right). \label{theta tensor}
\end{eqnarray}
To calculate \eqref{Wald mass int}, we take $\Sigma$ as the $(\theta, \phi)$ 2D surface and $\xi^\mu=(1,0,0,0)$ is the time-like Killing vector. The parameter $s$ defines the transition path from the original metric to the background metric, and after the substitutions $\mathcal{M}\rightarrow \mathcal{M}s$ and $a\rightarrow as$, it is not difficult to see that $s=0$ characterizes the background metric and $s=1$ the original metric. If we take into account the relations $R_{\mu\nu}=3k g_{\mu\nu}$ and $R=12k$, the second terms of \eqref{K tensor} and \eqref{theta tensor} are equal to zero.
\begin{equation}
    K^{tr}=\frac{16 \alpha  k \mathcal{M} s }{r^2}+\mathcal{O}\left(\frac{1}{r^3}\right), \; \; \Theta^r=\mathcal{O}\left(\frac{1}{r^3}\right) ds
\end{equation}
and the Wald mass is
\begin{eqnarray}
    M_\mathrm{W}=\frac{1}{8}\int\limits_0^1\int\limits^\pi_0 \lim_{r\rightarrow\infty} \left(\frac{\partial K^{tr}}{\partial s}d s+\frac{1}{2}K^{tr} \frac{4 a^2 s \cos^2{\theta}}{\rho^2}ds-\xi^t\Theta^{r}\right)r^2 \sin{\theta}d\theta=4\alpha_g k\mathcal{M}, \label{Wald mass}
\end{eqnarray}
If we choose the Killing vector $\xi^\mu=(0,0,0,-1)$ space-like, the expression \eqref{Wald mass int} gives the Weyl angular momentum $J_W=M_W a$. 
\subsection{Second law and phase transition}
According to the second law of thermodynamics, a BH, as a result of any external influence, if there is no external matter, its entropy should only increase. However, since in conformal Weyl gravity there is also an external background $\gamma$ and $\kappa$, we cannot say that it is in a vacuum. As we noted in the \ref{subsec: first law}, the entropy for a conformal Weyl BH consists of the sum of the entropy arising from the Einstein and Wald theories $S=S_\text{E}+S_\text{W}$. We test the entropy change separately using the perturbation method. First, we examine the entropy in terms of $\delta M_\text{Komar}$ and $\delta J_\text{Komar}$ according to Einstein's theory.
\begin{equation}
    \left(\frac{\partial S_\text{E}}{\partial\mathcal{M}}\right)_{J}=\frac{4 \left(\mathcal{M} \sqrt{\mathcal{M}^2-a^2}+\mathcal{M}^2\right)}{\sqrt{\mathcal{M}^2-a^2}}+2 k \left(\sqrt{\mathcal{M}^2-a^2}+\mathcal{M}\right)^4\frac{2 \mathcal{M}^2 \sqrt{\mathcal{M}^2-a^2}+2 a^2 \sqrt{\mathcal{M}^2-a^2}-3 a^2 \mathcal{M}+2 \mathcal{M}^3 }{\mathcal{M} \left(\mathcal{M}^2-a^2\right)^{3/2}} \label{dSE/dM}
\end{equation}
In \eqref{dSE/dM}, the first term  leads to the Hawking temperature in the case $k=0$ and is always positive, and in the condition 
\begin{equation}
    a^2<\left(\frac{5\sqrt{17}}{8} -\frac{13}{8}\right)\mathcal{M}^2\approx 0.952 \mathcal{M}^2
\end{equation}
the second term be positive, and since it is very close to the extreme case, expression \eqref{dSE/dM} can generally we can say to be positive. The change of Einstein's entropy by $\delta J_\text{Komar}$ is
\begin{equation}
    \left(\frac{\partial S_\text{E}}{\partial J}\right)_\mathcal{M}=-\frac{2 a}{\sqrt{\mathcal{M}^2-a^2}}-\frac{a k \left(\sqrt{\mathcal{M}^2-a^2}+\mathcal{M}\right)^4 \left(4 \mathcal{M}^2-4 a^2-\mathcal{M} \sqrt{\mathcal{M}^2-a^2}\right)}{\mathcal{M} \left(\mathcal{M}^2-a^2\right)^2}\label{dS/dj}
\end{equation}
and for always completely negative, the spin parameter must be $a^2<15 \mathcal{M}^2/16$. It is natural that the result is completely negative because we saw in Sec.\ref{sec: Extractable energy and PP} that as a result of the external influence, the BH loses its rotational energy $\delta J_\text{Komar}<0$. 

Now let's check the entropy change for the pure Weyl theory.
\begin{equation}
    \left(\frac{\partial S_\text{W}}{\partial M_\text{W}}\right)_{J_\text{W}}=-\frac{16 \pi ^2 \left(\sqrt{\mathcal{M}^2-a^2}+\mathcal{M}\right)^2 \left(\mathcal{M} \left(\sqrt{\mathcal{M}^2-a^2}+\mathcal{M}\right)+2 a^2\right)}{\mathcal{M}^2 \sqrt{\mathcal{M}^2-a^2}}<0,
\end{equation}
\begin{equation}
     \left(\frac{\partial S_\text{W}}{\partial J_\text{W}}\right)_{M_\text{W}}=\frac{24 \pi ^2 a \left(\sqrt{M^2-a^2}+M\right)^2}{M^2 \sqrt{M^2-a^2}}>0.
\end{equation}
Although the pure Weyl entropy alone exhibits $\partial S_W/\partial M_W < 0$, this does not violate the second law because the physical entropy is the $S_\text{total} = S_\text{E} + S_\text{W}$. As derived in \eqref{dSE/dM} and \eqref{dS/dj}, the Einstein part dominates for small $k$ and yields $\partial S_E/\partial M_E > 0$. Consequently, the total entropy increases with the total mass, and the second law is satisfied in the full Einstein–Weyl theory.

In BH thermodynamics, a phase transition is a change in the sign of the BH heat capacity $C_\text{J}$ , and for the Kerr BH the point $a_p=\sqrt{2\sqrt{3}-3}M\approx 0.681 M$ is the transition point, where the heat capacity has a second-order discontinuity. At $a<a_p$ the heat capacity is negative, and at $a>a_p$ it is positive. The heat capacity of BH is generally defined as follows:
\begin{equation}
    C_\text{J}=T \left(\frac{\partial S}{\partial T}\right)_J=T \left(\frac{\partial S_E}{\partial T}\right)_J+T \left(\frac{\partial S_W}{\partial T}\right)_J=C_E+C_W
\end{equation}
and the phase transition point can be determined by $\partial T/\partial \mathcal{M}=\partial \kappa_g/\partial \mathcal{M}=0$, which is a common point for the total heat capacity. Using \eqref{surface gravity}, we can approximate the phase transition point as follows
\begin{equation}
    a=a+\delta a\approx 0.681 \mathcal{M}+7.037 k \mathcal{M}^3\approx 
    M\left(0.681-1.022 \gamma M + 1.759 \gamma^2 M^2 + 7.037 \kappa M^2-31.668 \gamma\kappa M^3\right) \label{phase transition point}
\end{equation}
\eqref{phase transition point} shows that the thermodynamic critical point is shifted from its Kerr value by the conformal parameters $\gamma$ and $\kappa$. In the limit $\gamma,\kappa\rightarrow0$, the Kerr result $a/M=\sqrt{2\sqrt{3}-3}\simeq0.68125$ is recovered. The linear correction in $\gamma$ shifts the critical spin towards
smaller values, whereas the $\kappa$-dependent contribution shifts it towards larger values. The quadratic and mixed terms in the conformal parameters provide subleading corrections to this behaviour. Thus, the location of the heat-capacity singularity is not universal, but depends explicitly on the deformation of the underlying geometry.
This demonstrates that the conformal sector modifies the thermodynamic stability structure of the black hole by changing the spin at which the heat capacity diverges.
\begin{figure}
    \centering
    \includegraphics[scale=0.5]{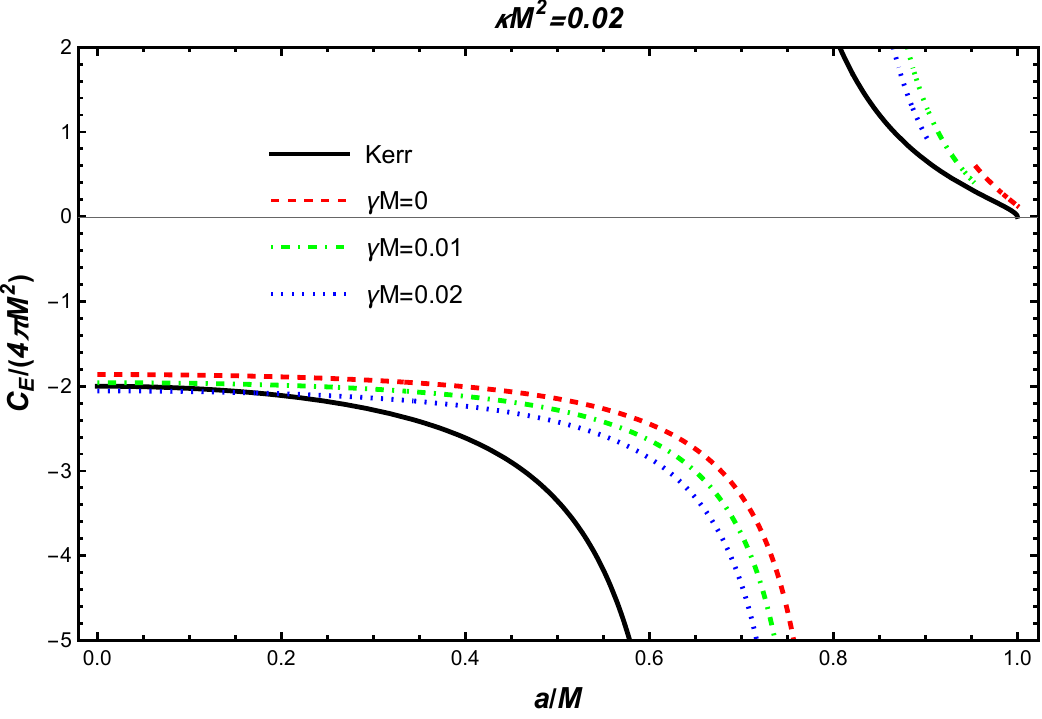}
    \includegraphics[scale=0.5]{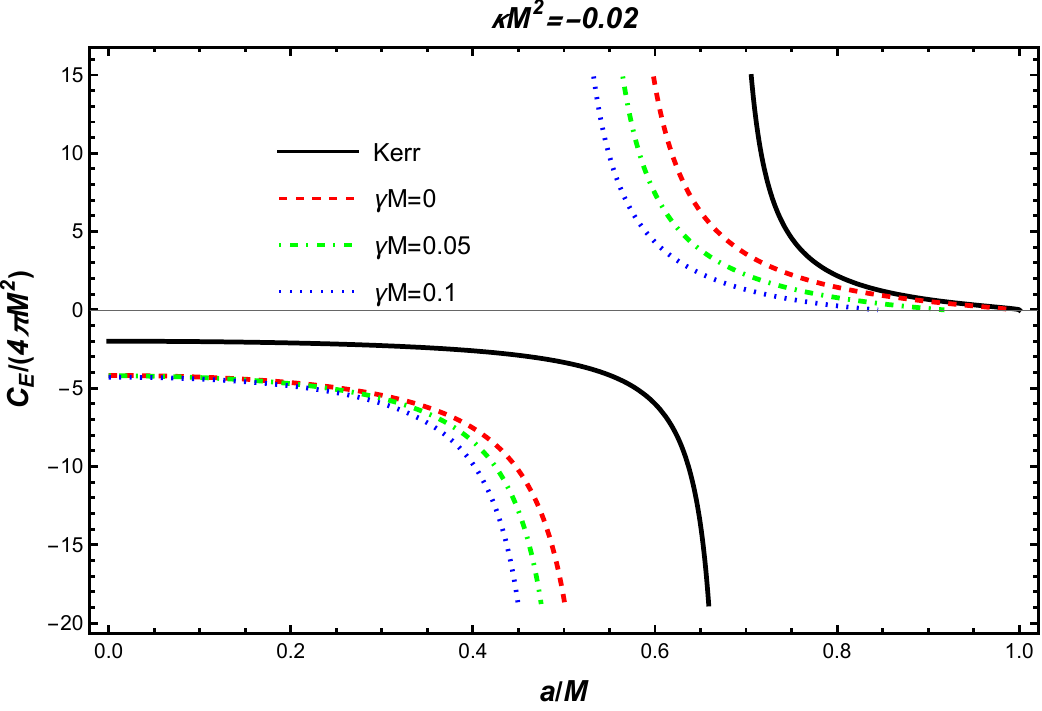}
    \includegraphics[scale=0.5]{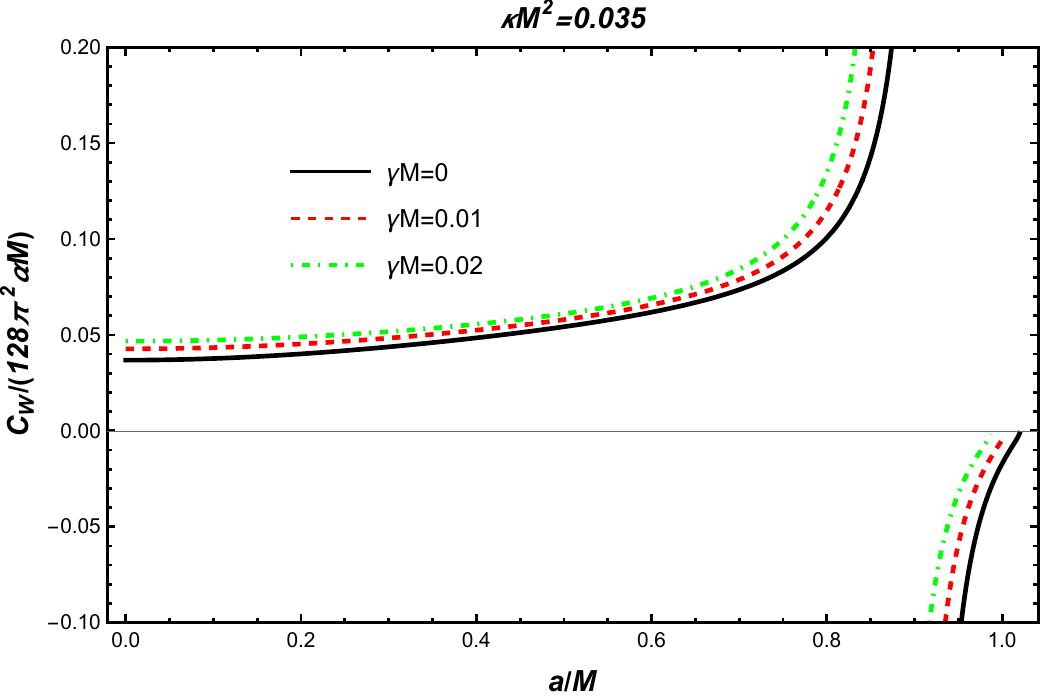}
    \includegraphics[scale=0.5]{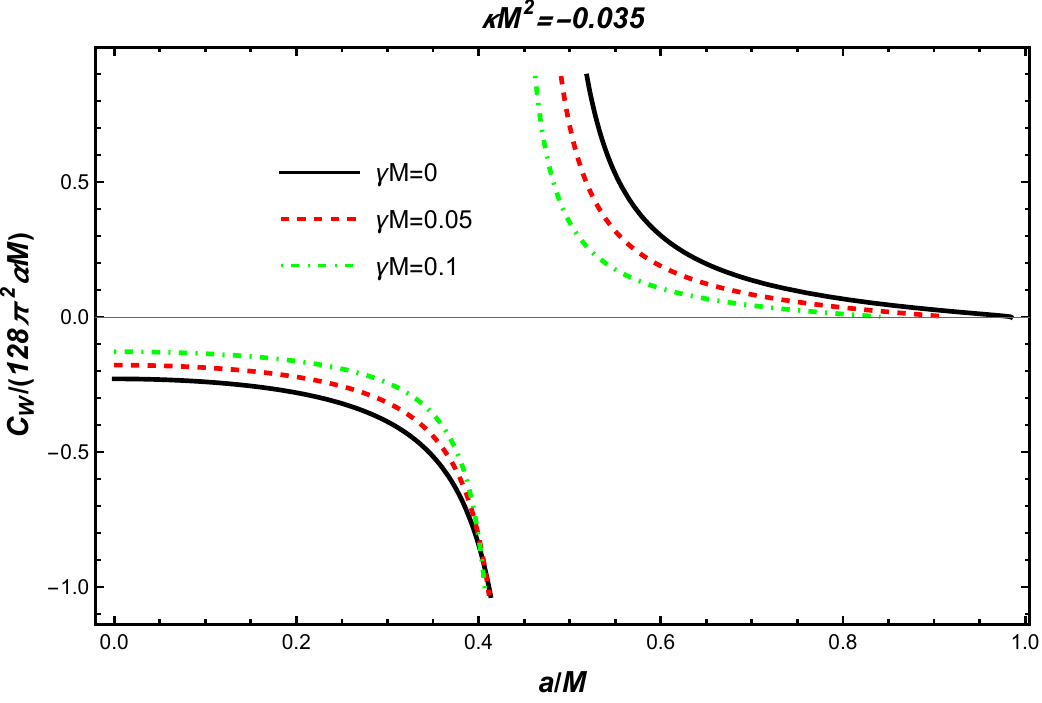}
    \caption{Dimensionless heat capacities $C_E/(4\pi M^2)$ (upper panels) and $C_W/(128\pi^2\alpha_g M)$ (lower panels) as functions of the 
    dimensionless spin parameter $a/M$ for selected values of the conformal 
    parameter $\gamma$, at fixed $\kappa M^2 = 0.02$ (left column) and 
    $\kappa M^2 = -0.02$ (right column) for the Einstein part, and at 
    $\kappa M^2 = 0.035$ (left) and $\kappa M^2 = -0.035$ (right) for 
    the Weyl part.}
    \label{fig:heat capacity}
\end{figure}

Fig.~\ref{fig:heat capacity} illustrates the dimensionless heat capacities $C_E/(4\pi M^2)$ and $C_W/(128\pi^2\alpha_g M)$ as functions of $a/M$ 
for representative values of $\gamma$ and both signs of $\kappa$ according to \eqref{outer horizon}.  Several features deserve attention. First, the divergence of $C_E$  identifies the phase transition point $a_p$, which shifts in accordance  with \eqref{phase transition point}: for $\kappa > 0$, a larger $\gamma$ moves $a_p$ to smaller spins, narrowing the thermodynamically stable window,  while for $\kappa < 0$ the opposite trend is observed. This sign-dependent competition between $\gamma$ and $\kappa$ reflects the nontrivial interplay 
already noted in the surface gravity (Fig.~\ref{fig:surface gravity}) and the 
extremality condition \eqref{eq:extreme case}. Second, the Weyl contribution $C_W$ is suppressed relative to $C_E$ by the small coupling $\alpha_g k$, yet it undergoes its own sign change near $a_p$, demonstrating that the higher-derivative sector responds to the same instability. Consequently, the total heat capacity $C_J = C_E + C_W$ inherits a phase structure that is qualitatively similar to the Kerr case  but is quantitatively modified by both conformal parameters independently, providing clear thermodynamic signatures of conformal Weyl gravity.

\subsection{Third law}
In the original PP Sec.\ref{BH slow down}, we saw that an infinite number of massive particles would be needed to stop the BH from spinning and retain some of its mass. This is consistent with the third law of thermodynamics, which states that the surface area or entropy of the BH cannot be reduced to zero by a finite number of processes. According to the third law of classical thermodynamics, when the temperature of a system drops to absolute zero, the entropy should also tend to zero. However, in BHs, even when $T\rightarrow 0$, there is a zero state, in which $S\neq0$. It can be seen from equation \eqref{surface gravity} that the state $T=0$ corresponds to the extreme cases, and using \eqref{eq:extreme case}, we can calculate the entropy approximately as follows
\begin{align}
    S\Big|_{T\rightarrow 0} & \rightarrow  \pi  \mathcal{M}^2 \left(2+5 k \mathcal{M}^2\right)+32 \pi ^2 \alpha  \left(2-k \mathcal{M}^2\right)\nonumber\\ &\approx  2 \pi  \left(32 \pi  \alpha +M^2\right)+\pi  \kappa  \left(5 M^4-32 \pi  \alpha  M^2\right)-6\pi \gamma M^3+\left(\frac{23 \pi  M^4}{4}-8 \pi ^2 \alpha  M^2\right)\gamma^2
\end{align}
The finite entropy at extremality does not violate the third law because the heat capacity $C_\text{J}$ vanishes as $T\rightarrow0$ (since $C_\text{J}\sim T^2$ near extremality). Moreover, the impossibility of reaching the exact extreme state by any finite physical process (due to the infinite number of required particle injections) ensures that the third law holds in its usual formulation for BH thermodynamics. Hence, conformal Weyl gravity respects the third law. 

\section{Conclusions}\label{sec:conclusion}
In this work, we have systematically investigated the energy extraction mechanisms and thermodynamic properties of rotating black holes in conformal Weyl gravity. We have considered the Kerr-like black hole solution characterized by two dimensionful parameters, $\gamma$ and $\kappa$, which respectively modify the asymptotic behaviour of the gravitational field on galactic and cosmological scales. Unlike previous studies that often combine these parameters into a single effective parameter $k$, we have explicitly analysed their individual contributions to the horizon structure, particle dynamics, and thermodynamic quantities.

We derived the horizon structure and provided both perturbative and exact expressions for the event horizons. We found that $\gamma$ and $\kappa$ play opposite roles in determining the extremal configuration: a positive $\kappa$ increases both the horizon radius and the maximal allowed spin, whereas a positive $\gamma$ reduces them. This competition between the conformal parameters enriches the phase space of black hole solutions compared to the Kerr case and demonstrates that the two parameters cannot be treated as a single effective coupling without losing essential physical information. The original Penrose process was analysed in detail. We computed the extractable energy and the efficiency of energy extraction, showing that the conformal corrections reduce the maximum extractable energy relative to the Kerr geometry. In the extremal limit, the efficiency approaches $\epsilon \simeq 0.112$ when $k\mathcal{M}^2=2/27$, which is slightly below the Kerr value. We also derived the black hole spin-down law, which reveals that the mass required to stop the black hole rotation is modified by the conformal parameter $k$, while the qualitative feature that an infinite number of particles is needed remains unchanged.

Turning to thermodynamics, we verified that the four laws of black hole thermodynamics hold in conformal Weyl gravity. Using the Wald entropy formalism, we obtained the Weyl contribution to the entropy $S_\text{W}$. The surface gravity and Hawking temperature were computed, and we explicitly demonstrated that the conformal parameters shift the phase transition point of the heat capacity. Notably, the sign of $\kappa$ plays a crucial role: for $\kappa>0$, increasing $\gamma$ lowers the critical spin, while for $\kappa<0$ the opposite trend occurs. This sign-dependent interplay between $\gamma$ and $\kappa$ provides a clear thermodynamic signature of conformal Weyl gravity.

Overall, our findings demonstrate that conformal Weyl gravity introduces small but physically meaningful modifications to the Kerr black hole thermodynamics and energy extraction efficiency. These corrections are controlled by the dimensionless combinations $\gamma M$ and $\kappa M^2$, which are extremely small for astrophysical black holes, but they become relevant in regimes where conformal gravity is expected to play a role, such as in the early universe or near galactic centres. The explicit separation of the effects of $\gamma$ and $\kappa$ highlights the predictive power of conformal gravity and provides a clear target for observational tests of modified gravity theories.

\section*{Data Availability Statement}
The data generated and analyzed in this study are not publicly available due to their size and computational nature. Also, it can be obtained from the corresponding author upon reasonable request.
\bibliographystyle{unsrt}
\bibliography{reference}

\end{document}